\documentclass[%
 reprint,
superscriptaddress,
 amsmath,amssymb,
 aps,
 prb,
floatfix,
]{revtex4-2}
\usepackage[T1]{fontenc}
\usepackage{graphicx}
\usepackage{dcolumn}
\usepackage{bm}
\usepackage{soul}
\soulregister{\cite}7
\soulregister{\ref}7
\usepackage{color,xcolor}
\usepackage{subfigure}
\usepackage[colorlinks,
			linkcolor=blue,
            anchorcolor=blue,
            citecolor=blue,
            urlcolor=blue]{hyperref}
\begin{document}

\preprint{APS}

\title{Acceptor-based qubit in silicon with tunable strain}

\author{Shihang Zhang}
\author{Yu He}
 \email{hey6@sustech.edu.cn}
\author{Peihao Huang}  
 \email{huangph@sustech.edu.cn}
 \affiliation{Shenzhen Institute for Quantum Science and Engineering, Southern University of Science and Technology, Shenzhen 518055, China}%
 \affiliation{International Quantum Academy, Shenzhen 518048, China}
 \affiliation{Guangdong Provincial Key Laboratory of Quantum Science and Engineering, Southern University of Science and Technology, Shenzhen 518055, China}

\date{\today}

\begin{abstract}
Long coherence time and compatibility with semiconductor fabrication make spin qubits in silicon an attractive platform for quantum computing. In recent years, hole spin qubits are being developed as they have the advantages of weak coupling to nuclear spin noise and strong spin-orbit coupling (SOC), in constructing high-fidelity quantum gates. However, there are relatively few studies on the hole spin qubits in a single acceptor, which requires only low density of the metallic gates. In particular, the investigation of flexible tunability using controllable strain for fault-tolerant quantum gates of acceptor-based qubits is still lacking. Here, we study the tunability of electric dipole spin resonance (EDSR) of acceptor-based hole spin qubits with controllable strain. The flexible tunability of LH-HH splitting and spin-hole coupling (SHC) with the two kinds of strain can avoid high electric field at the "sweet spot", and the operation performance of the acceptor qubits could be optimized. Longer relaxation time or stronger EDSR coupling at low electric field can be obtained. Moreover, with asymmetric strain, two "sweet spots" are induced and may merge together, and form a second-order "sweet spot". As a result, the quality factor $Q$ can reach $10^{4}$ for single-qubit operation, with high tolerance for the electric field variation. Furthermore, the two-qubit operation of acceptor qubits based on dipole-dipole interaction is discussed for high-fidelity two-qubit gates. The tunability of spin qubit properties in acceptor via strain could provide promising routes for spin-based quantum computing.
\end{abstract}

\maketitle

\section{Introduction}\label{section:intro}

Spin-based qubit in silicon is an important candidate platform for quantum computation. The original framework of the spin qubit in semiconductor is based on the electron spin and nuclear spin \cite{Loss1998,Kane1998}. In the past decades, electron spin qubits in gate-defined quantum dots (QDs) are well developed \cite{Zwanenburg2013,Huang2014,Tahan2014,Petta2005,Zajac2018}. In particular, high-fidelity ($>99\%$) single-qubit and two-qubit gates are realized \cite{Noiri2022,Xue2022}. In contrast, qubit based on hole spin is established and developed \cite{Bulaev2007,Ruskov2013,Salfi2016PRL,Ono2017,Hendrickx2020N,Scappucci2020,Hendrickx2021,Zwerver2022,Camenzind2022} rather late and less attention has been paid. However, strong intrinsic spin-orbit coupling of holes allows all-electrical manipulation of qubits without additional design\cite{Bulaev2007}. Moreover, the suppressed coupling between a hole spin and nuclear spins in host material also reduces pure spin dephasing \cite{Bulaev2005,Fischer2008}. For hole spin qubits, four-qubit device and long coherence time has been achieved \cite{Hendrickx2021,Zwerver2022}. Except for spin qubit in gate-defined QDs, dopant-based spin qubit is another potential option by using scanning tunneling microscope (STM) lithography \cite{Broome2017,Broome2018,Campbell2022} or ion implantation \cite{Pla2013}. And compared to the gate-defined QDs, single-atom devices provide more a steady environment, which induces a long relaxation time of qubits \cite{Hsueh2014,Watson2017}. And particles are confined deeply by the potential of the atom nucleus, reducing the density of gates \cite{Hile2015}. High-fidelity two-qubit gates based on donor atoms have also been realized \cite{Madzik2022}. Although both hole spin qubits and single-donor qubits are developed intensively in experiments, single-acceptor-based hole spin qubits are still less studied. Due to strong spin-orbit coupling and low density of gates, acceptor-based qubits could have advantages for fast quantum gates, long coherence time, and high scalability. It is shown that the operation of acceptor-based qubits may be realized via electric dipole spin resonance (EDSR) \cite{Salfi2016PRL}. And a long coherence time of 10 ms of acceptor spins is demonstrated \cite{Kobayashi2021}.

Energy levels of spin states in the acceptor can be influenced by the interaction with the electric field, magnetic field, and strain \cite{Golding2003,Salfi2020}. The acceptor system is more complex than donors due to these interactions and the spin-3/2 system \cite{Winkler2004,Culcer2006,Winkler2008}. The hole spin states split into two degenerate spin states, called heavy hole (HH) spin states and light hole (LH) spin states (in host material). Experimentally, the readout of acceptor qubits in Si:B device shows the special energy levels of the spin-3/2 system and long relaxation time \cite{Heijden2014t,Heijden2018,Kobayashi2021}. For now, the experiment on operations of acceptor-based qubits is lacking. Traditionally, the qubit operation is via electron spin resonance based on the oscillating magnetic field, which is hard to generate locally and enhance. To avoid these difficulties, all-electrical manipulation of spin qubits is expected. For electron spin qubit or hole spin qubit in QD, the all-electrical qubit operation is realized by electric dipole spin resonance (EDSR) based on spin-orbit coupling (SOC), engineered by magnetic field gradient or hyperfine interaction (flip-flop) \cite{Golovach2006,Tokura2006,Tosi2017}. High-speed qubit operations based on EDSR are the key to high-fidelity qubit gates. Similarly, the manipulation of the acceptor-based qubit states via EDSR is proposed based on coupling between LH states and HH states with opposite spin polarization, called spin-hole coupling (SHC) \cite{Salfi2016PRL,Salfi2016Nano}. 

However, SOC or SHC also makes qubits sensitive to charge noise, which induces decoherence of qubits \cite{Huang2014,Bermeister2014}. Fortunately, the sensitivity of qubits to the charge noise can be reduced even during operation with controlling pulse or energy level engineering \cite{Hu2006,Horibe2015,Martins2016,Reed2016}. For the acceptor qubit, there is an operation point immune to the first-order electrical noise, named "sweet spot", which was predicted theoretically \cite{Salfi2016PRL,Salfi2016Nano}. In the acceptor system, the appearance of the sweet spot is a combined effect of LH-HH splitting and SHC. And manipulation of acceptor-based qubits also depends on these mechanisms. Thus, the LH-HH splitting and SHC are critical underlying physics for the spin qubit operations. In the previous study, both of them are tuned solely by the vertical electric field \cite{Salfi2016Nano}. Consequently, the operation performance of the qubits is limited since the electric field can not control the two quantities (i.e.  LH-HH splitting and SHC) independently. For example, in previous work, to access the sweet spot too high electric field may be required for the system, and the operation performance can not be improved easily \cite{Salfi2016PRL}, where only a constant strain was introduced to make light hole states ground states \cite{Salfi2016PRL,Kobayashi2021}. Moreover, the tolerance to the electric field noise may be small. In this work, we show these problems can be solved by introducing tunable strain into the system. Furthermore, the tunability of the strain and the new SHC mechanism induced by strain are considered in our work. 

In this work, we study the electric manipulation of an acceptor spin qubit in the presence of tunable strain. The operation performance of acceptor-based qubits can be optimized by strain engineering. Firstly, strain can adjust the LH-HH splitting $\Delta_{LH}$. When the qubits operate at the sweet spot, the main decoherence comes from the relaxation due to phonon. Larger LH-HH splitting induces longer relaxation time, which improves the coherence of acceptor qubits. And importing SHC from asymmetric strain, the EDSR coupling may be enhanced. Consequently, both single-qubit and two-qubit operation rates are higher. With proper strain, two sweet spots for the electric field appear at low electric field. The two sweet spots can merge together where a second-order sweet spot immune to the charge noise appears. We find regions where the spin qubit has high quality factor and high tolerance to the electric field at the same time. As a result, with tunable strain, high-fidelity single-qubit and two-qubit gates beyond the fault-tolerant threshold could be constructed based on all-electrical manipulations. The two-qubit gates can be realized with long-range coupling between two qubits. In conclusion, we demonstrate a feasible scheme based on acceptor spin qubits for a large-scale fault-tolerant quantum computer.

This paper is developed as follows: In Sec. \ref{section: model}, the model and Hamiltonian of the system with strain are introduced. Based on that, the qubit definition and EDSR of the acceptor qubit are detailed in Sec. \ref{section: qubit}. Decoherence of the qubit is also introduced. In Sec. \ref{section: strain}, the results with effects of the strain are discussed. And two-qubit operation via electric dipole-dipole interaction is introduced in Sec. \ref{section: two}. In the last section, the conclusion of this work is given. An outlook on future research on this topic is mentioned.

\begin{figure*}[htbp]
\centering
\includegraphics[width=0.8\linewidth]{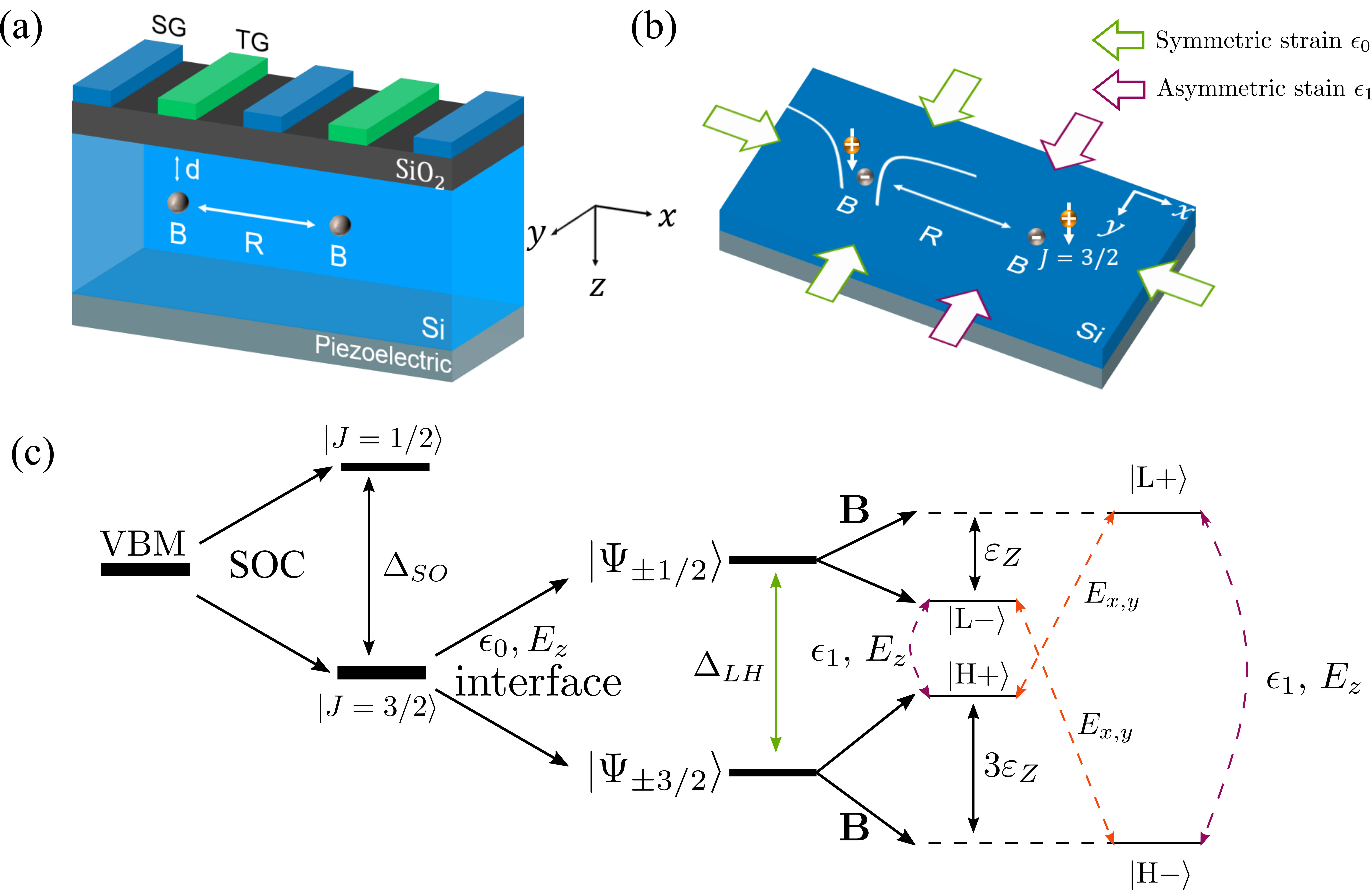}
\caption{The schematic diagram of the model. (a) The schematic diagram of the device. Boron acceptors are implanted in silicon near the Si/SiO$_{2}$ interface with the depth of $d$. They are separated by the distance of R. Top gates (TG) generate the vertical electric field. Side gates (SG) are used to apply in-plane electric fields. The gray layer can be piezoelectric materials, which stretch the silicon layer to produce strain. (b) The schematic diagram of the silicon layer. The green edge arrows indicate the static strain induced by the mismatch of lattice parameters. And the purple edge arrows are asymmetric strain in $\hat{y}$ direction. Notice that the tensile strain on $\hat{z}$-axis is not shown. The spin-3/2 states of the holes (yellow) bound to the acceptor atoms (silver) are spin qubits. (c) Schematic of energy levels of hole spin states. Due to the spin-orbit coupling, states on the six-fold degenerate valence band maximum (VBM) are split into a two-fold degenerate state named split-off band and a four-fold degenerate state. Then the four-fold degenerate states can be divided into the heavy hole and light hole states, by strain or electric field. The magnetic field further splits the 2-fold degeneracies of the spin states of the heavy hole and light hole. In this work, the heavy hole spin state is always grounded. The qubit states are spin-hole mixed states due to interaction with asymmetric strain and $T_{d}$-interaction with electric field $E_{z}$. The LH-HH transition due to the in-plane electric field plays an important role in EDSR.}
\label{fig:device0}
\end{figure*}

\section{Model}\label{section: model}

As shown in Fig. \ref{fig:device0}, boron atoms are placed near the interface in the silicon with a depth of $d$. In principle, the boron atom can be placed by using ion implantation or STM lithography \cite{Fuechsle2010,Morello2010,Heijden2014t,Heijden2018,Campbell2022}. On the top of the device, gate electrodes are used to manipulate the acceptor qubit. A magnetic field is applied along the $\hat{z}$-axis perpendicular to the surface (see Fig. \ref{fig:device0} (a)). Strain ($\epsilon=\epsilon_{0}+\sum_{i=x,y}\epsilon_{ii1}$) is introduced to the system in two ways: (i). Growing the silicon heterostructure on the $\mathrm{Si}_{x}\mathrm{Ge}_{1-x}$ or using thermal expansion of different materials \cite{Asaad2020,Kobayashi2021}. The strain created in this way is static and 'symmetric' ($\epsilon_{xx}=\epsilon_{yy}=-(C_{11}/2C_{12})\epsilon_{zz}$) \cite{AbadilloUriel2017}. $C_{11}$ and $C_{12}$ are the elastic stiffness constants for the strain-stress tensor. In this work, the strain created in this way is labeled as $\epsilon_{0}=\epsilon_{xx0}+\epsilon_{yy0}-\epsilon_{zz0}$, named symmetric strain. (ii). Applying piezoelectric material to produce strain in a certain direction \cite{Ranjan2021,Neill2021}. In this case, asymmetric in-plane strain can be obtained ($\epsilon_{xx}\neq \epsilon_{yy}$). And it can be tuned by electric field \cite{Neill2021}. In this work, strain induced in this way is labeled as $\epsilon_{ii1} $ $(i=x,y)$. An asymmetric strain ($\epsilon_{xx}\neq \epsilon_{yy}$) can induce new coupling mechanism like SHC independent on the vertical electric field \cite{Bir19631}. To tune the SHC without changing LH-HH splitting, strain should have opposite deformation: $\epsilon_{xx1}=-\epsilon_{yy1}$, called asymmetric strain. For simplicity, asymmetric strain is denoted as: $\epsilon_{1}=\epsilon_{xx1}-\epsilon_{yy1}=2\epsilon_{xx1}$. The system Hamiltonian based on the device is:
\begin{equation}
H=H_{\mathrm{Lut}}+H_{\epsilon}+H_{Z}+H_{E}.
\end{equation}
$H_{\mathrm{Lut}}$ is Luttinger Hamiltonian \cite{Luttinger1955}. It's a 4x4 matrix describing the heavy hole and light hole in bulk. The split-off band is ignored here as their energy levels are well separated from the states of interest. $H_{\epsilon}$ is Bir-Pikus Hamiltonian describing the interaction with strain, detailed in Supplementary \cite{supp}:
\begin{equation}
\begin{aligned}
H_{\epsilon}=a'Tr[\epsilon]+b'((J_{x}^{2}-\frac{5}{4}I)\epsilon_{xx}+c.p.)\\+(2d'/\sqrt{3})(\{J_{x},J_{y}\}\epsilon_{xy}+c.p.),
\end{aligned} 
\end{equation}
where $\epsilon_{ij}$ is strain, c.p. means cyclic permutations, the $a'$, $b'$ and $d'$ are deformation potential, $\{J_{x},J_{y}\}=1/2(J_{x}J_{y}+J_{y}J_{x})$ is anti-commutator of spin-3/2 operator. $H_{Z}$ is the Zeeman Hamiltonian describing the interaction with the magnetic field:
\begin{equation}
H_{Z}=\mu_{B}[g_{1}(J_{x}B_{x}+c.p.)+g_{2}(J_{x}^{3}B_{x}+c.p.)].
\end{equation} 
Only the $\hat{z}$ direction and its linear term is considered (the g factor $g_{1}=1.07>>g_{2}$) \cite{Baldereschi1973a}. Here, $+\hat{z}$-axis is pointing down towards silicon. $H_{E}=H_{C}+H_{if}+H_{gate}+H_{T_{d}}$ includes Hamiltonian describing interaction with electric field: $H_{C}=e^{2}/4\pi \epsilon_{s} r$ is Coulomb potential, $r$ is position of hole relative to the nuclei of the acceptor and $\epsilon_{s}$ is static dielectric constant of semiconductor. $H_{if}=U_{0}\Theta(-z)$ is interaction with the interface potential. And $H_{gate}=e\mathbf{E\cdot r}$ represents the interaction with the interface gate field. These interface terms play an important role due to the large transition between the light hole and heavy hole states, called LH-HH coupling. $H_{T_{d}}=2pE_{x}\{J_{y},J_{z}\}/\sqrt{3}+c.p.$ is interaction with the electric field due to the tetrahedral ($T_{d}$) symmetry of acceptor in silicon \cite{Bir19632}. The c.p. is cyclic permutation and $\{A,B\}=(AB+BA)/2$. In the equation, $p=e\int_{0}^{a}f^{*}(r)rf(r)\textrm{d}r$ is effective dipole moment, where $a$ is the lattice constant and $f(r)$ is radial bound hole envelope function. The $J_{i}$ are matrices of the spin-3/2 for $i=x,y,z$.

To construct the matrix of the Hamiltonian, we define the spin states of the heavy hole and light hole as basis states. As shown in Fig. \ref{fig:device0}. (c), the six-fold degenerate state of the valence band in silicon is split into a two-fold degenerate state named split-off band, and a topmost four-fold degenerate state due to the SOC \cite{Luttinger1956}. Near the interface between the silicon and the dielectric layer, the topmost four-fold degenerate state of the valence band with $J=3/2$ is separated into two doubly degenerate states. They can be named heavy hole (HH) states ($|H\pm\rangle=|m_{J}=\pm3/2\rangle$) and light hole (LH) states ($|L\pm\rangle=|m_{J}=\pm1/2\rangle$). Spin-orbit coupling mixes states with $\Delta L=0,\pm 2$. Consequently, the wavefunctions (in bulk) of the orbital ground states are linear combinations of states with $L=0$ and $L=2$ \cite{Baldereschi1973a}. And the orbital excited states are also linear combinations of states with different $L$.

The Hamiltonian in the subspace of the orbital ground states $\{|H+\rangle,|H-\rangle,|L+\rangle,|L-\rangle\}$ is:
\begin{widetext}
\begin{equation}\label{Eq:H}
H=\left(\begin{array}{lccc}
\varepsilon_{\mathrm{H}+} & 0 & -i p E_{+}+\alpha E_{-} & t^{*} \\
0 & \varepsilon_{\mathrm{H}-} & t & -i p E_{-}-\alpha E_{+}  \\
i p E_{-}+\alpha E_{+}  & t^{*} & \varepsilon_{\mathrm{L}+} & 0 \\
t & i p E_{+}-\alpha E_{-}  & 0 & \varepsilon_{\mathrm{L}-}
\end{array}\right),
\end{equation}
\end{widetext}
where $\varepsilon_{i}$ represents the energy of state $i$ ($i=|H\pm\rangle, |L\pm\rangle$): $\varepsilon_{\mathrm{H}\pm}=\varepsilon_{H}\pm (3/2)\varepsilon_{Z}$, $\varepsilon_{\mathrm{L}\pm}=\varepsilon_{L}\pm (1/2)\varepsilon_{Z}$ ($\varepsilon_{Z}=g\mu_{B}B$). In this work, we let $g=1.07$ for holes in silicon \cite{Kopf1992}. The off-diagonal terms (SHC) $t= \mathrm{i}  p E_{z} + \frac{\sqrt{3}}{2}b'(\epsilon_{xx}-\epsilon_{yy}) $ mix hole and spin states provided by not only $T_{d}$ interaction with $E_{z}$, but also interaction with asymmetric strain. These terms mix $|H+\rangle$ ($|H-\rangle $) and $|L-\rangle$ ($|H+\rangle $) defining the qubit states, which will be discussed later. The $E_{\pm}=E_{x}\pm iE_{y}=Ee^{\pm i\theta_{E}}$ are related to in-plane electric field. $\theta_{E}$ defines the direction of the in-plane electric field. Terms related to $E_{\pm}$, include two parts: One is the $T_{d}$ interaction with the in-plane electric field with $p$. Another is interaction with interface potential and electric field with coefficient $\alpha$, which is obtained by projecting the orbital first-excited states onto the ground states via Schrieffer–Wolff transformation \cite{Winkler,Salfi2016Nano}. Both of them couple the HH and LH states with spin states unchanged and produce an LH-HH coupling, which will be a key mechanism to drive spin qubits in this paper. Therefore, the time-dependent in-plane electric field can be utilized to drive qubits. Combining the effect and the SHC terms, the manipulation of acceptor qubit by a process similar to electric dipole spin resonance (EDSR) can be achieved \cite{Golovach2006}, which will be introduced in the following section. The splitting between the heavy hole and light hole state is \cite{Salfi2016PRL}:
\begin{equation}
\Delta_{LH}=\varepsilon_{L}-\varepsilon_{H}=\Delta_{if}+\Delta(E_{z})+\Delta_{\epsilon},
\end{equation}
where $\Delta_{if}$ is from the interface potential \cite{Mol2015,AbadilloUriel2016}, $\Delta(E_{z})$ depends on the gate electric field and $\Delta_{\epsilon}=b'(\epsilon_{xx}+\epsilon_{yy}-\epsilon_{zz})=b'(\epsilon_{0}+\epsilon_{xx1}+\epsilon_{yy1})$. In the model, tuning of strain can change the LH-HH splitting $\Delta_{LH}$ and SHC $t$ 
individually, which plays an important role in qubit operation. In the following section, qubit definition and operation will be introduced.

\section{Qubit and operation}\label{section: qubit}
This section is the basis for the following section. \ref{section: strain}. In this section, the qubit, its operation, and decoherence with strain will be discussed. The effect of LH-HH splitting and spin-hole coupling on operation performance will be highlighted, for a better understanding of the results in the next section.

\subsection{Qubit definition}\label{sec:qubit0}

 \begin{figure*}[!htbp]
 \centering
 \includegraphics[width=1\textwidth]{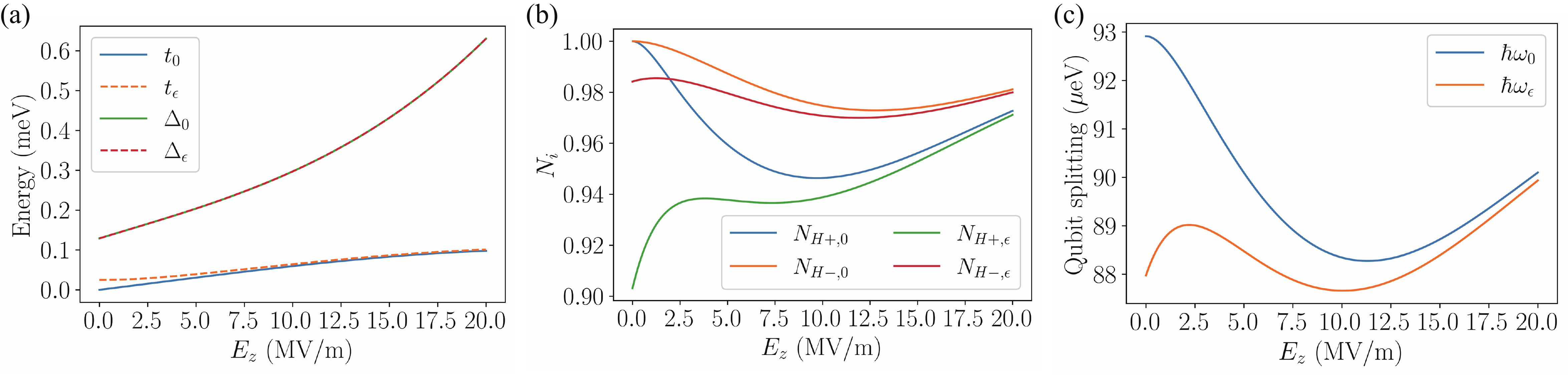}%
 \caption{Explanation of appearance of sweet spots without or with strain ($\epsilon_{xx}=-\epsilon_{yy}=-10^{-5}$) at $B=0.5$ T, $d=4.6$ nm. (a) LH-HH splitting and SHC without strain ($\Delta_{0}$, $t_{0}$) and with strain ($\Delta_{\epsilon}$, $t_{\epsilon}$) are plotted as function of vertical electric field $E_{z}$. In particular, SHC appears without the electric field in the presence of asymmetric strain. (b) Proportion of $|H\pm\rangle$ in the qubit states $|g\pm\rangle$ without strain ($N_{g\pm,0}$) and with strain ($N_{g\pm,\epsilon}$) are plotted as function of vertical electric field $E_{z}$. (c) Qubit splitting without ($\hbar\omega_{0}$) and with strain ($\hbar\omega_{\epsilon}$) are plotted as a function of the vertical electric field $E_{z}$ with strain. Corresponding with the variation of proportions of states, qubit splitting increases at the low electric field. Then it turns to decrease and is the same trend along the situation without strain.}
 \label{fig:sweet spot}
 \end{figure*} 
 
From Eq. (\ref{Eq:H}), we can define the acceptor qubit by diagonalization of static parts of the Hamiltonian. Then, the manipulation of the qubit can be induced by the time-dependent in-plane electric field, which will be included after the diagonalization. To make sure the HH spin states are the ground states, we assumed that $\Delta_{LH}>2\varepsilon_{Z}$. Then, after diagonalization, the eigenvalues are:
\begin{equation}\label{Eq:qubit-sy}
\begin{aligned}
\lambda_{g+}&= \frac{1}{2}\left[\varepsilon_{\mathrm{H}+}+\varepsilon_{\mathrm{L}-}-\tilde{\Delta}_{-+}\right],  \\
\lambda_{g-}&=\frac{1}{2}\left[\varepsilon_{\mathrm{H}-}+\varepsilon_{\mathrm{L}+}-\tilde{\Delta}_{+-}\right], \\
\lambda_{\mathrm{e}+}&=\frac{1}{2}\left[\varepsilon_{\mathrm{H}-}+\varepsilon_{\mathrm{L}+}+\tilde{\Delta}_{+-}\right], \\
\lambda_{\mathrm{e}-}&=\frac{1}{2}\left[\varepsilon_{\mathrm{H}+}+\varepsilon_{\mathrm{L}-}+\tilde{\Delta}_{-+}\right],
\end{aligned}
\end{equation}
where $\tilde{\Delta}_{\mp \pm}=\sqrt{\Delta_{\mp \pm}^{2}+4t^{2}}$, $\Delta_{\mp \pm}=\varepsilon_{\mathrm{L}\mp}-\varepsilon_{\mathrm{H}\pm}$ and $t=\sqrt{(3/4)b^{'2}\epsilon_{1}^{2}+p^{2}E^{2}}$ is SHC. And corresponding eigenvectors:
\begin{equation}\label{Eq:states}
\begin{split}
\left|g+\right\rangle &=\frac{1}{N_{1}}\left(\begin{array}{c}
-\frac{\Delta_{-+}+\tilde{\Delta}_{-+}}{2} \\
0 \\
0 \\
t
\end{array}\right)=\frac{1}{N_{1}}\left(\begin{array}{c}
-a\\
0 \\
0 \\
t
\end{array}\right),\\
\left|g-\right\rangle &=\frac{1}{N_{2}}\left(\begin{array}{c}
0 \\
\frac{\Delta_{+-}+\tilde{\Delta}_{+-}}{2}\\
-t^{*} \\
0
\end{array}\right)=\frac{1}{N_{2}}\left(\begin{array}{c}
0\\
c \\
-t^{*} \\
0
\end{array}\right),\\
\left|\mathrm{e}+\right\rangle &=\frac{1}{N_{2}}\left(\begin{array}{c}
0 \\
-t\\
-\frac{\Delta_{+-}+\tilde{\Delta}_{+-}}{2}\\
0
\end{array}\right)=\frac{1}{N_{2}}\left(\begin{array}{c}
0\\
-t \\
-c \\
0
\end{array}\right),\\
\left|\mathrm{e}-\right\rangle &=\frac{1}{N_{1}}\left(\begin{array}{c}
t^{*}\\
0 \\
0 \\
\frac{\Delta_{-+}+\tilde{\Delta}_{-+}}{2}
\end{array}\right)=\frac{1}{N_{1}}\left(\begin{array}{c}
t^{*}\\
0 \\
0 \\
a
\end{array}\right),
\end{split}
\end{equation}
where $N_{i}$ (i=1, 2) is normalization coefficients:
\begin{equation}
\begin{aligned}
N_{1}=\sqrt{|t|^2+a^2},N_{2}=\sqrt{|t|^2+c^2}.
\end{aligned}
\end{equation}
As mentioned in the last section, $T_{d}$ interaction with the vertical electric field and asymmetric strain mix the hole and spin states. We define our qubits on the lowest two states: \{$|0\rangle =|g-\rangle, |1\rangle = |g+\rangle$\}. The qubit states are mixing of the spin-hole states illustrated in Fig. \ref{fig:device0}. (c). They are mostly the HH spin states. The qubit splitting can be obtained:
\begin{equation}\label{Eq:hw}
\hbar\omega=\lambda_{g+}-\lambda_{g-}=2\varepsilon_{Z}-\tilde{\Delta}_{-+}+\tilde{\Delta}_{+-}.
\end{equation}

The qubit splitting depends directly on the LH-HH splitting $\Delta_{LH}$, SHC $t$, and the magnetic field $B_{z}$. The derivative $\partial\hbar\omega/\partial E_{z}=0$ defines the so-called sweet spot. Operation of qubit at the sweet spot can be immune to electrical noise. Moreover, when $\partial\hbar\omega/\partial E_{z}=0$ and $\partial^{2}\hbar\omega/\partial E_{z}^{2}=0$, qubit is immune to the second order electrical noise, defines the second-order sweet spot. To explain the existence of the sweet spot, the compositions of qubit states should be emphasized. Acceptor qubit state $|g+\rangle$ $(|g-\rangle)$ is a mixture of $|H+\rangle$ $(|H-\rangle)$ and $|L-\rangle$ $(|L+\rangle)$. Therefore, the energy splitting of eigenstates could be changed by tuning their mixing proportion. The proportions of states in the qubit states are determined by the relative strength of $\Delta_{LH}$ and $t=|t_{\pm}|$. With proper strength of $\Delta_{LH}$ and $t$, the proportion of states will not be varied with the electric field. Mathematically, the sweet spot appears at the extreme point of qubit splitting. An example is given in Fig. \ref{fig:sweet spot} (c). For $d=4.6$ nm without strain (blue line), there is a minimum around $E_{z}=12$ MV/m for qubit splitting, which is the sweet spot. The mechanism behind this is the competition between the SHC $t_{0}$ and LH-HH splitting $\Delta_{0}$. As shown in Fig. \ref{fig:sweet spot} (a), the dominant factor is varying in different regions. When the electric field $E_{z}<12$ MV/m, the mixing of states is enhanced due to the increasing SHC $t_{0}$. However, when $E_{z}>12$ MV/m, the mixing of states is reduced. That is because where the LH-HH splitting $\Delta_{0}$ increases significantly and plays a dominant role. Thus, the sweet spot appears around $E_{z}=12$ MV/m, which is the extreme point of qubit splitting.

The following section will show that another sweet spot could exist and can merge with the first sweet spot in presence of asymmetric strain. We compare the situation with and without strain ($\epsilon_{xx}=-\epsilon_{yy}=-10^{-3}$ \% ), shown in Fig. \ref{fig:sweet spot}. At $E_{z}=0$ point, despite the electric field disappearing, the SHC $t$ still exists due to the asymmetric strain. Thus, at $E_{z}=0$, the states are already mixed. In Fig. \ref{fig:sweet spot} (b), the components of $|H+(-)\rangle$ in $|g+(-)\rangle$, defined as $N_{H+(-)}=|a(c)/N_{1(2)}|^{2}$, is not equal to 1 in the absence of electric field $E_{z}$. As mentioned above, the qubit splitting is determined by SHC $t$ and LH-HH splitting $\Delta$. In Fig. \ref{fig:sweet spot}. (c), with asymmetric strain $\epsilon_{1}$ and within $0<E_{z}<2.5$ MV/m, the qubit splitting $\hbar\omega_{\epsilon}$ increases as the electric field $E_{z}$ increases. In the region, the qubit splitting $\hbar\omega_{\epsilon}$ is mainly affected by the variation of LH-HH splitting $\Delta_{LH}$. That is because as $E_{z}$ increases, $\Delta_{LH}$ is increased while the variation of the SHC $t_{\epsilon}$ is negligible, where the strain-induced SHC dominates over the electric-field-induced SHC. However, when the electric field $2.5<E_{z}<10$ MV/m, the qubit splitting $\hbar\omega_{\epsilon}$ starts decreasing because the SHC $t_{\epsilon}$ is increasing, where the electric-field-induced SHC becomes dominant. In addition to the sweet spot discussed in the last paragraph, a new sweet spot appears at around $E_{z}=2.5$ MV/m, in the presence of the asymmetric strain. The two sweet spots are getting close to each other for larger SHC induced by the strain. Later on, we show that the two sweet spots can merge together and form a second-order sweet spot with tunable strain.

\subsection{Operation}\label{sec: operation}

\begin{figure*}[htbp]
	\centering
    \includegraphics[width=1\textwidth]{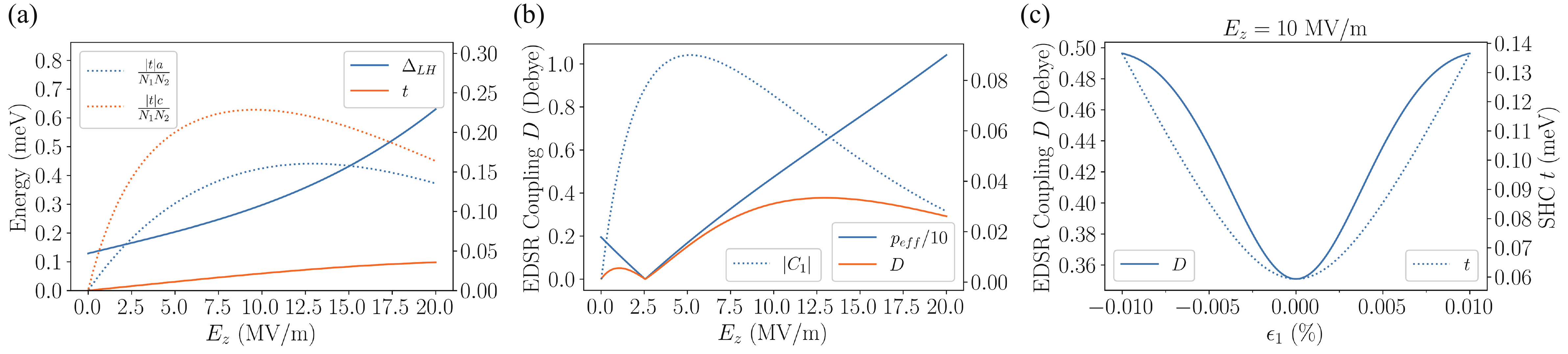}%
	\caption{Explanation of variation of the EDSR coupling when $d=4.6$ nm, B=0.5 T. To simplify the process, we assume that there is no strain in (a) and (b). (a) The variation of the ratio factor of states as $E_{z}$ increases. (b) The EDSR coupling $D$ and its factors $p_{eff}$ and $C_{1}$. (c). The enhancement of EDSR coupling $D$ by asymmetric strain with the vertical electric field $E_{z}=7$ MV/m. The asymmetric strain ($\epsilon_{xx}=-\epsilon_{yy}$) is assumed to introduce SHC without changing LH-HH splitting.} 
	\label{fig:edsr}
\end{figure*}

The qubit operation of the acceptor qubit is induced by utilizing the LH-HH transition, providing state mixing. As mentioned above, the qubit states are the spin-hole mixed states. The $|H\mp\rangle$ in $|g\mp\rangle$ can be coupled to $|L\mp\rangle$ in $|g\pm\rangle$ by LH-HH transition, which can be used to realize single-qubit operation. Specifically, an applied AC electric field can modulate the LH-HH transition. And thus, the LH-HH transition provides coherent driving of the acceptor qubit. Single qubit operation is obtained by writing these terms on the qubit basis:
\begin{widetext}
\begin{equation}
H_{E_{\parallel}}'=\left(\begin{array}{lccc}
0 & C_{1}(\alpha E_{-}-ipE_{+}) & C_{2}(\alpha E_{-}-ipE_{+}) &0 \\
C_{1}^{*}(\alpha E_{+}+ipE_{-}) & 0 & 0& -C_{2}(\alpha E_{+}+ipE_{-})  \\
C_{2}(\alpha E_{+}+ipE_{-}) & 0 & 0& C_{1}(\alpha E_{+}+ipE_{-})\\
0 & -C_{2}(\alpha E_{-}-ipE_{+})& C_{1}^{*}(\alpha E_{-}-ipE_{+}) & 0
\end{array}\right),
\end{equation}
\end{widetext}
where $C_{1}=t^{*}(a-c)/N_{1}N_{2}$, $C_{2}=(|t|^{2}+ac)/N_{1}N_{2}$. The EDSR dipole moment is
\begin{equation}
D=p_{eff}|C_{1}|= |t|p_{eff} \frac{|a-c|}{N_{1}N_{2}},
\end{equation}
where $p_{eff}=\sqrt{\alpha^{2}+p^{2}+2\alpha p\sin(2\theta_{E})}$ is the effective LH-HH transition. In this work, we assume that $E_{x}=E_{y}$, i.e. $p_{eff}=|\alpha+p|$. Once the electric dipole moment $D$ is obtained, the acceptor qubit operation can be modulated by the electric dipole transition $DE_{ac}$, where $E_{ac}$ is an in-plane alternating electric field. The Rabi frequency can be calculated: $\omega_{R}=DE_{ac}/h$. The strength of $D$ is determined by $p_{eff}$ and $C_{1}$. The $p_{eff}$ includes $p$ and $\alpha$, which depends on the electric field $E_{z}$. However, the interface-induced $\alpha$ is much larger than $p$ induced by electric field $E_{z}$ \cite{Salfi2016Nano}. Remarkably, $C_{1}$ comes from difference between $a$ and $c$. Here, the $\frac{|t|a(c)}{N_{1}N_{2}}$ represents the transition between $|H+(-)\rangle$ and $|L+(-)\rangle$ in qubit states, which depends on the magnetic field $B$, LH-HH splitting $\Delta_{LH}$ and SHC $t$. The influence of $\Delta_{LH}$ and $t$ is plotted in Fig. \ref{fig:edsr}. (a). The dependences of $a$ and $c$ on the LH-HH splitting $\Delta_{LH}$ and spin-hole coupling $t$ are different. The $a$ is more sensitive to the change due to the smaller splitting between $|H+\rangle$ and $|L+\rangle$. As shown in Fig. \ref{fig:edsr} (b), the trend of the $p_{eff}$ and $|C_{1}|$ is opposite in two regions. Thus, EDSR coupling $D$, which is the product of them, has two peaks in each region. In Fig. \ref{fig:edsr} (b), there is a dip around the vertical electric field of 2.5 MV/m. This is mainly because $p_{eff}$ approaches zero around 2.5 MV/m. Moreover, the larger magnetic field can enhance the difference between a and c. That is because larger Zeeman splitting reduces $a$ (less $|H+\rangle$ in $|g+\rangle$), and enhances $c$ (more $|H-\rangle$ in $|g-\rangle$). In conclusion, $D$ can be enhanced by the magnetic field, the LH-HH transition $p_{eff}$ and SHC $t$, and be lowered by LH-HH splitting $\Delta_{LH}$.

For example, SHC $t$ induced by asymmetric strain can enhance the EDSR coupling, shown in Fig. \ref{fig:edsr}. (c). However, the enhancement is maximized around $|\epsilon_{xx}|=0.01$ \%. That is because the coupling between $|H+\rangle$ and $|L-\rangle$ in $|g+\rangle$ stops increasing when the SHC is large enough. Meanwhile, the coupling in between $|H-\rangle$ and $|L+\rangle$ in $|g-\rangle$ is still increasing. Then, the difference $C_{1}$ is reduced.

\subsection{Decoherence}
Coherence time $T_{2}$ is as crucial as operation speed for quantum computation. The performance of the qubits can be estimated by the quality factor $Q=\omega_{R}T_{2}/2\pi$, which is the operation times of a full rotation before the qubit states decohere. In silicon, the dephasing of hole spin qubit due to the hyperfine interaction can be reduced by isotopic purification \cite{Witzel2007,Tyryshkin2012,Chekhovich2012}. Thus, the pure dephasing of the acceptor qubit is mainly induced by the charge noise. The pure dephasing rate $1/T_{\phi}=\delta E^{2}\tau/(2\hbar^{2})$ \cite{Bermeister2014} is calculated in supplementary materials \cite{supp}, considering the energy fluctuation $\delta E$ caused by the extra electric field to the second order. The electric field due to the defect is assumed as 3380 V/m \cite{Salfi2016Nano}. The energy fluctuation depends on the derivative $\partial \hbar\omega/\partial E_{z}$ and $\partial^{2}\hbar\omega/\partial E_{z}^{2}$. As mentioned above, for the acceptor-based qubit, there exist sweet spots where the qubit splitting is insensitive to the variation of the electric field. Thus, operation at the sweet spot can reduce the dephasing related to the electrical noise. From our model, the pure dephasing is greatly suppressed at the sweet spot, where the relaxation becomes the main decoherence source of the qubit. This is further illustrated in the following section.

\begin{figure*}[!htbp]
\centering
\includegraphics[width=1\textwidth]{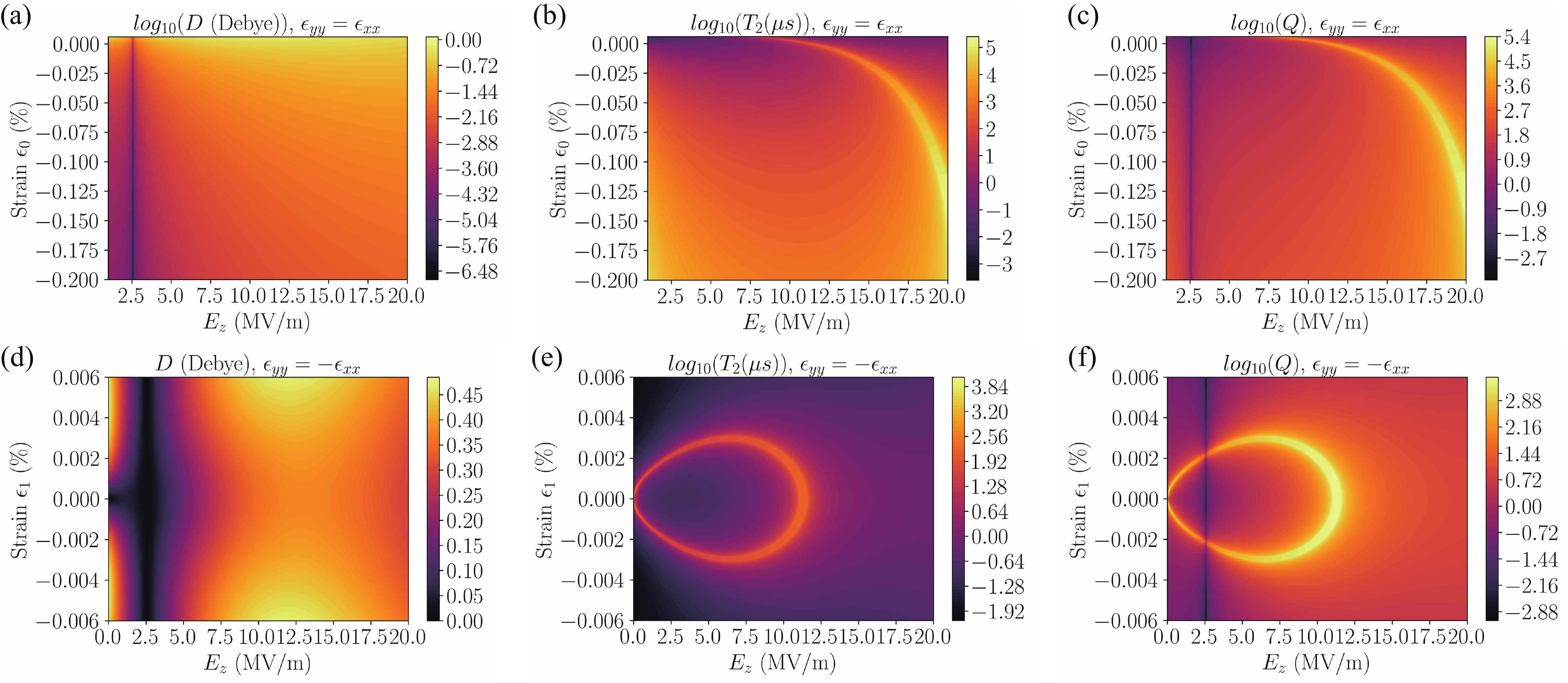}%
\caption{Effect of symmetric and asymmetric strain. $B=0.5$ T, $E_{ac}=10^{4}$ V/m, $d=4.6$ nm. (a), (b), (c) are all plotted as function of symmetric strain $\epsilon_{0}$ and vertical electric field $E_{z}$. (a) The EDSR coupling $D$. (b) Decoherence time $T_{2}$ depends on the relaxation time $T_{1}$ and the pure dephasing time $T_{\phi}$. At the sweet spot, the relaxation due to phonon leads the decoherence. (c) The quality factor $Q$. In general, the quality factor $Q$ is enhanced from hundreds to $10^{4}$. The performance of the qubit operation is better with the strain. However, observing (b), the sweet spot requires a larger electric field with higher static strain, which can be solved by SHC induced by asymmetric strain, seeing (d), (e), (f). For these figures, the spin-hole coupling is enhanced by the difference between the strain in $\hat{x}$ and $\hat{y}$ direction. And they are all plotted as a function of asymmetric strain $\epsilon_{1}$ and vertical electric field $E_{z}$. (d) The EDSR coupling $D$. (e) The decoherence time $T_{2}$. There are two sweet spots for proper strain. (f) The quality factor $Q$. The quality factor is almost unchanged and slightly smaller with asymmetric strain.}
\label{fig:static strain}
\end{figure*}

The main source of relaxation of hole spin qubits is spin-phonon interaction via deformation potential \cite{Ehrenreich1956,Srivastava1990}. The spin relaxation of the qubits can be calculated as
\begin{equation}
 1/T_{1}=\frac{(\hbar \omega)^{3}}{20 \hbar^{4} \pi \rho}(C_{1})^{2}\left[2d^{'2}\left(\frac{2}{3 v_{l}^{5}}+\frac{1}{v_{t}^{5}}\right)\right],
\end{equation}
where $\rho=2330$ kg/m$^{3}$ is the mass density, $d'=-3.7$ eV is the deformation potential \cite{Neubrand}, $v_{l}=899$ m/s and $v_{t}=1.7v_{l}$ are the longitudinal and transverse sound velocities in silicon. The relaxation rate has a quadratic dependence on the $C_{1}$. By contrast, EDSR coupling $D$ depends on $C_{1}$ linearly. As mentioned above, $C_{1}$ is determined by SHC $t$ and LH-HH splitting $\Delta_{LH}$. That means relaxation of qubit will be more sensitive to the variation of parameters of the system, than the EDSR operation rate. For example, as shown in Sec. IV of supplementary material \cite{supp}, $T_{1}$ is enhanced four orders of magnitude when $E_{z}=10$ MV/m, compared to the case in the absence of strain. While the EDSR coupling is enhanced two orders of magnitude with the same condition in Fig. \ref{fig:static strain}. (a). As a result, the quality factor (the number of operations within coherence time) is enhanced when the larger LH-HH splitting is induced by strain.


\section{Effect of strain}\label{section: strain}

This section shows the effect of strain and is divided into two parts. In the first part, we discuss qubit operation performance by tuning LH-HH splitting $\Delta_{LH}$ and SHC $t$ with strain separately. And then, the LH-HH splitting and SHC will be tuned simultaneously. The optimal points for qubit operation are found by plotting them as a function of strain and electric field.

In this work, we assume: The depth of the acceptor $d=4.6$ nm, the magnetic field $B=0.5$ T, and the strength of the in-plane electric field $E_{ac}=10^{4}$ V/m. For silicon, the Bir-Pikus deformation potential $b'=-1.42$ eV and $d'=-3.7$ eV\cite{Neubrand}. The LH-HH splitting $\Delta_{LH}$, effective dipole moment $p$, and interface-induced spin-hole coupling $\alpha$ depend on the wavefunctions of hole spin states. They could be calculated numerically \cite{Baldereschi1973a,AbadilloUriel2016}. Given that they have been already calculated, we take them from a previous work \cite{Salfi2016Nano}. In the first part, the symmetric strain is set as $\epsilon_{0}\in [-0.1,0.003]$ $\%$, to ensure heavy hole states are the ground states. And the asymmetric strain is set as $\epsilon_{1}\in [-0.003,0.003]$ $\%$.

\subsection{Effect of 'symmetric' and 'asymmetric' strain}\label{sec: static}
To show the effect of tuning of LH-HH splitting on the qubit operation, a 'symmetric' strain is considered, which means $\epsilon_{xx}=\epsilon_{yy}$. For simplicity, the strain is denoted as $\epsilon_{0}=\epsilon_{xx}+\epsilon_{yy}=2\epsilon_{xx}$. Thus the LH-HH splitting induced by the strain (mostly compressed) is $\Delta_{\epsilon}=b'\epsilon_{0}$. Note that $b'$ is negative, $\Delta_{\epsilon}$ grows as strain decreases. The EDSR coupling $D$, decoherence time $T_{2}$ and quality factor $Q$ are plotted in (a), (b) and (c) of Fig. \ref{fig:static strain}. EDSR coupling is reduced due to the larger LH-HH splitting in the presence of strain. In Fig. \ref{fig:static strain}. (b), the coherence time $T_{2}$ is dominated by the dephasing. However, at sweet spots, the decoherence is mainly due to relaxation. The relaxation time $T_{1}$ is reduced with more negative symmetric strain. From Section. \ref{section: model}, when the LH-HH splitting increases (strain $\epsilon_{0}$ decreases), the proportion number $|C_{1}|$ increases. Consequently, the EDSR coupling decreases and relaxation time increases, inferred from their dependences $|C_{1}|$. Moreover, the variation of relaxation time is square to that of EDSR coupling. In general, the quality factor $Q$ raises with more negative strain, as recognized in Fig. \ref{fig:static strain}. (c). And from Fig. \ref{fig:static strain}. (b), the sweet spot requires a higher vertical electric field $E_{z}$ with stronger symmetric strain $\epsilon_{0}$. Therefore, there is a trade-off between the quality factor and the required vertical electric field and strain. This is because the splitting $\Delta_{LH}$ increases as $\epsilon_{0}$ becomes more negative, then the strength of the SHC required to obtain the sweet spot is larger. For now, the only source of SHC is interaction with the vertical electric field $E_{z}$. In short, qubit operation with high quality factor $Q$ can be obtained with large symmetric strain, despite requiring a high vertical electric field $E_{z}$.

By applying the strain in x and y direction with opposite deformation ($\epsilon_{xx1}=-\epsilon_{yy1}$, that is asymmetric strain), the spin-hole coupling, $t=\sqrt{3}b'\epsilon_{1}/2+ipE_{z}$ is tuned independently by asymmetric strain, while LH-HH splitting by strain is fixed. The EDSR coupling $D$, decoherence time $T_{2}$ and quality factor $Q$ are plotted in (d), (e) and (f) of Fig. \ref{fig:static strain}. Certainly, the effect of SHC on the qubit operation is symmetric for positive and negative $\epsilon_{1}$. In Fig. \ref{fig:static strain} (d), in general, EDSR coupling is enhanced. And the EDSR coupling is weak around $E_{z}=2.5$ MV/m. The enhancement in EDSR coupling is due to the larger $C_{1}$ with extra SHC from asymmetric strain. For weak electric fields, the electric field-induced SHC is weak. Thus, the enhancement due to strain is more significant. However, when the electric field is large, there is no significant influence of SHC induced by strain on EDSR coupling. The dip appears despite enhanced $C_{1}$ due to the weak $p_{eff}$. The other impact of tuning of SHC is on the decoherence of the qubit. In Fig. \ref{fig:static strain} (f), with the tuning of strain, high values of $Q$ can be found in a wide region of the electric field $E_{z}\in [0,11]$ MV/m. The quality factor $Q$ is slightly reduced, compared to the case without asymmetric strain $\epsilon_{1}$. That is because $|C_{1}|$ is enhanced by SHC induced by asymmetric strain. And $p_{eff}$ is weaker at low electric fields. However, the electric field at the first-order sweet spot can be tuned by asymmetric strain. And the appearance of the second-order sweet spot, induced by asymmetric strain, protects the qubit in a wider electric field region. Moreover, the quality factor $Q$ is not reduced too much at the second-order sweet spot.

\begin{figure*}[bt]
\centering
\includegraphics[width=1\textwidth]{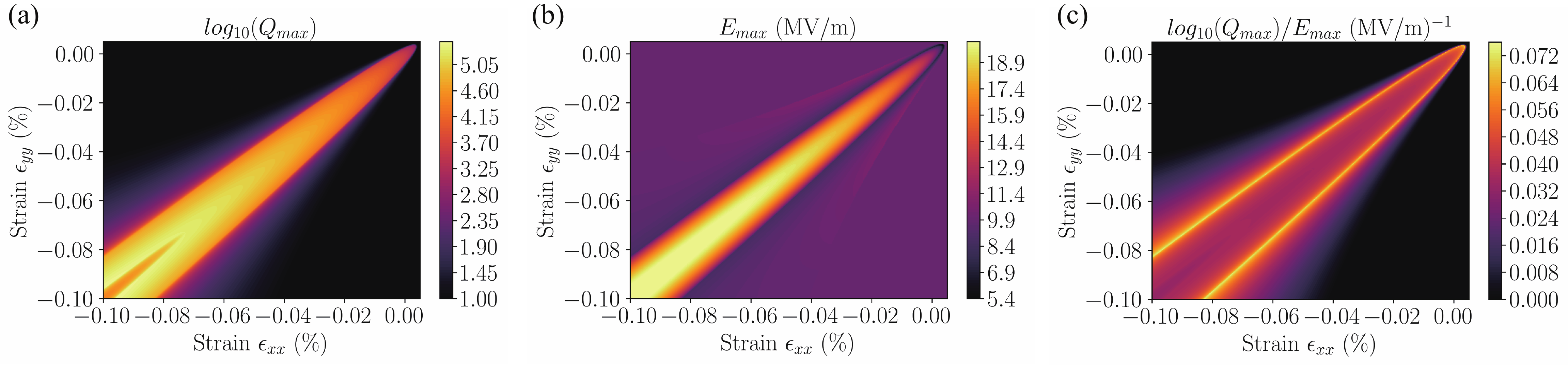}%
\caption{The combined effect of the strain. $B=0.5$ T, $E_{ac}=10^{4}$ V/m, $d=4.6$ nm. The window of the vertical electric field is set as $[1,20]$ MV/m. The figures are plotted as a function of strain $\epsilon_{xx}$ and $\epsilon_{yy}$. (a) The largest quality factor $Q_{max}$ is obtained by varying with strain $\epsilon_{xx}$ and $\epsilon_{yy}$. For $Q_{max}<10$, let $log_{10}(Q_{max})=1$. (b) The vertical electric field $E_{z}$ corresponds to $Q_{max}$, is denoted as $E_{max}$. The electric field $E_{max}$ becomes weaker away from the diagonal. For $Q_{max}<10$, let $E_{max}=10$ MV/m. (c) The efficiency of $Q_{max}$ on electric field $E_{z}$.}
\label{fig:strain}
\end{figure*}

\begin{figure*}[!htbp]
\centering
\includegraphics[width=1\textwidth]{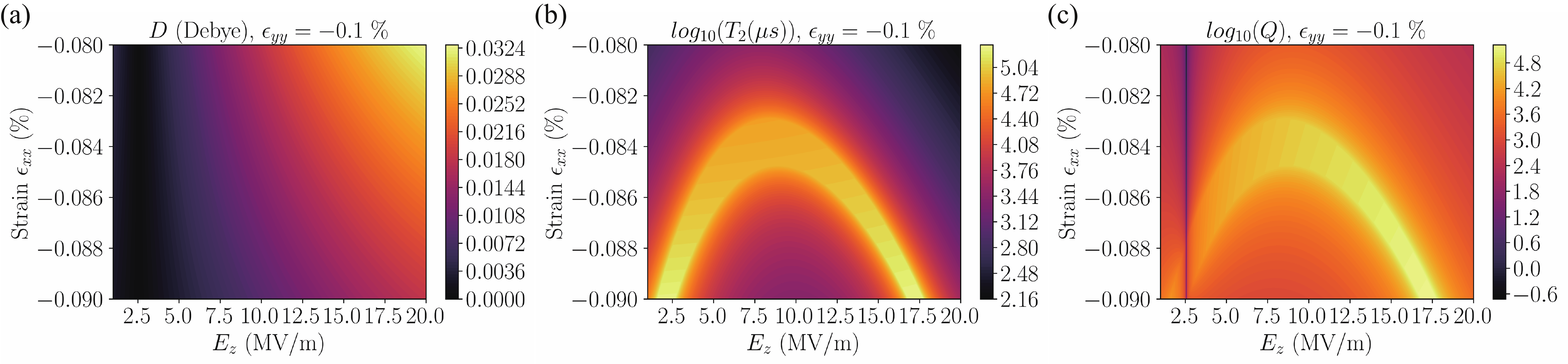}%
\caption{Qubit performance around an optimal region for operation. $B=0.5$ T, $E_{ac}=10^{4}$ V/m, $d=4.6$ nm. The asymmetric strain is set as $\epsilon_{yy}=-0.1$ \%. And the figures are plotted as a function of the vertical electric field $E_{z}$ and strain $\epsilon_{xx}$. (a). The EDSR coupling $D$. (b). The decoherence time $T_{2}$. Two sweet spots appear. (c). The quality factor $Q$.}
\label{fig:combined strain}
\end{figure*} 

\subsection{Combined effect of strain}\label{sec: combined}

According to the results shown in the last subsection, a 'symmetric' strain brings high quality factor requiring a large vertical electric field. An 'asymmetric' strain tunes the location of sweet spots without changing $Q$ too much. In this part, we show that qubit operation can be optimized by tuning the LH-HH splitting and SHC with strain. In Fig. \ref{fig:strain}, quantities are plotted as function of $\epsilon_{xx}$ and $\epsilon_{yy}$. In Fig. \ref{fig:strain}. (a), $Q_{max}$ is the maximum of quality factor $Q$ relative to the vertical electric field $E_{z}\in[1,20]$ MV/m, with a given strain. High $Q_{max}$ is concentrated around the diagonal area, called high-Q area ($Q_{max}>10^{3}$) below. The high $Q_{max}$ is obtained by operating qubits at the sweet spot. The high-Q area is at the first sweet spot. That is because $Q$ is larger for larger LH-HH splitting $\Delta_{LH}$ at the first sweet spot. Out of the area, there is no sweet spot for the strain condition, or the electric field corresponds to it out of the range [1,20] MV/m. In Fig. \ref{fig:strain}. (b), the electric field $E_{max}$ corresponding to the $Q_{max}$ is plotted in Fig. \ref{fig:strain}. (b). In the high-Q area as shown in Fig. \ref{fig:strain}. (a), the $E_{max}$ becomes smaller with larger difference between $\epsilon_{xx}$ and $\epsilon_{yy}$. And in the upper right corner, $Q_{max}$ can exceed 1000 with weak strain and electric field. 

\begin{figure*}[bt]
\centering
\includegraphics[width=1\textwidth]{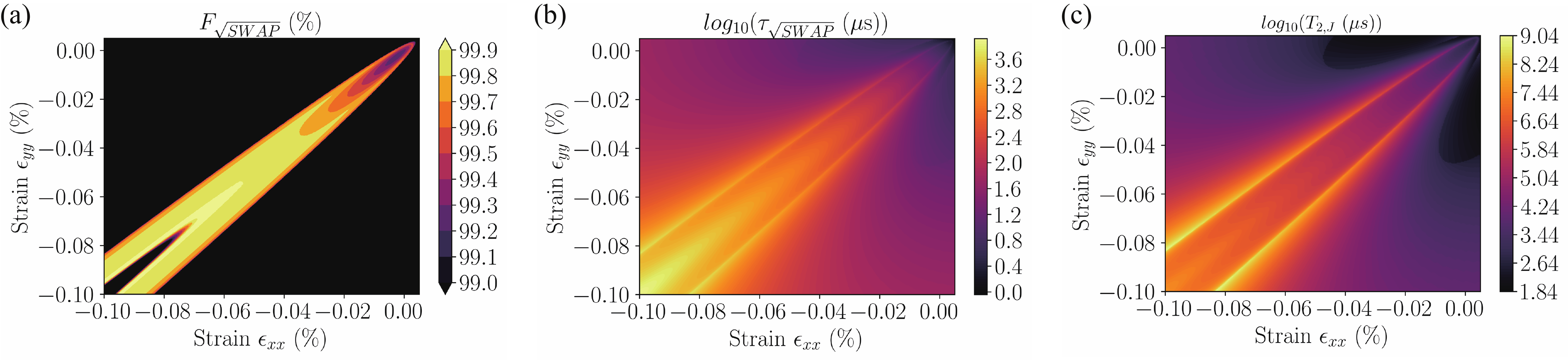}%
\caption{Operation performance for $\sqrt{\text{SWAP}}$ gate. $B=0.3$ T, $E_{ac}=10^{4}$ V/m, $d=4.6$ nm and $R_{12}=25$ nm. The window of the vertical electric field is set as $[1,20]$ MV/m. (a) The fidelity $F_{\sqrt{\text{SWAP}}}$, corresponding to the $Q_{max}$ for $\sqrt{\text{SWAP}}$ gate. (b) The SWAP operation time $\tau_{\sqrt{\text{SWAP}}}$ corresponds to $Q_{\sqrt{\text{SWAP}}}$. (c) The dephasing time $log_{10}(T_{\phi,J})$ due to the variation of $J_{dd}$ corresponding to $Q_{\sqrt{\text{SWAP}}}$. The dephasing is weaker than relaxation at the sweet spot. The $\sqrt{\text{SWAP}}$ operation time $\tau_{\sqrt{\text{SWAP}}}$ corresponds to $Q_{\sqrt{\text{SWAP}}}$.}
\label{fig:SWAP}
\end{figure*} 

To clarify the trade-off between the quality factor and electric field, $log_{10}(Q_{max})/E_{max}$ is plotted in Fig. \ref{fig:strain}. (c). There is a highly efficient line preserving high quality factors with low electric fields. Actually, it is reasonable. The electric field $E_{max}$ on the bright line corresponds to the 'sweet region', where the two sweet spots merge together. And similarly, to find the optimal point for tolerance to the electric field. The full width at half maximum (FWHM) of $E_{max}$, defined as $E_{tol}$, is plotted in Sec. V of the supplementary material \cite{supp}. We find $E_{tol}>1000$ V/m in most of the high-Q area, which is important for a robust qubit operation. In conclusion, an optimal region for qubit operation is found.

We choose a region in Fig. \ref{fig:strain} that assuming $\epsilon_{yy}=-0.1$ $\%$ and $\epsilon_{xx} \in [-0.09,-0.08]$ $\%$. The results are plotted in Fig. \ref{fig:combined strain}. In Fig. \ref{fig:combined strain}. (a), the EDSR coupling is slightly reduced around the sweet spot, compared with that without strain. In Fig. \ref{fig:combined strain}. (b), due to the enhancement of LH-HH splitting, relaxation is weaker in general. And with the aid of asymmetric strain, sweet spots can be obtained with a lower electric field. Around $\epsilon_{0}=-0.085$ \%, there is a second-order sweet spot with a high tolerance for the vertical electric field $E_{z}$. The decoherence of the qubit is determined by both relaxation and dephasing. Thus, around the sweet spots, the quality factor $Q$ can exceed $10^{4}$ in a wide region. For a small electric field, $Q$ still can exceed $10^{3}$.

\section{Two-qubit entanglement: dipole-dipole interaction}\label{section: two}
Long-range entanglement of spin qubits is crucial for the realization of scalable quantum circuits. Compared with short-range entanglement via Heisenberg exchange \cite{Joecker2021,madzik2021}, the long-range scheme mitigates the difficulties of fabrication by reducing the density of gates \cite{Hile2015}. For acceptor-based spin qubits, long-range entanglement can be realized by electric dipole-dipole interaction \cite{Salfi2016PRL,AbadilloUriel2018}. The interaction is based on the Coulomb interaction of qubits with each other. A $\sqrt{\text{SWAP}}$ gate can be constructed with coupling $J_{dd}\approx \frac{D^{2}}{4\pi\epsilon R_{12}^{3}}$, see Sec. VI of the supplementary material \cite{supp}. In the equation, $D$ is EDSR coupling, $\epsilon$ is permittivity for silicon, and $R_{12}$ is the distance between the acceptors. Operation time of $\sqrt{\text{SWAP}}$ gate is: $\tau_{\sqrt{\text{SWAP}}}=h/4J_{dd}$. The fidelity $F_{\sqrt{\text{SWAP}}}$ corresponding to $Q_{max}$ of $\sqrt{\text{SWAP}}$ gate is plotted as function of the strain $\epsilon_{xx}$ and $\epsilon_{yy}$. The distance between the acceptor atoms $R_{12}$ is 25 nm. And the magnetic field $B$ is 0.3 T, which is smaller than that used for single-qubit gates. A smaller magnetic field can enhance the fidelity of the qubit operation, see Sec. II of the supplementary material \cite{supp}. As a result, the fidelity $F_{\sqrt{\text{SWAP}}}$ can exceed the fault-tolerance threshold \cite{Fowler2012}. In particular, the fidelity can even reach up to 99.9 \%, when the strain $-0.07 \%<\epsilon_{xx} \approx \epsilon_{yy} <-0.05 \%$.

The region with high quality factor for $\sqrt{\text{SWAP}}$ is similar to the high-Q area for single-qubit operation. The enhancement effect of symmetric strain on $Q_{\sqrt{\text{SWAP}}}$ is less substantial for the $\sqrt{\text{SWAP}}$ gate. That is because the strength of the coupling $J_{dd}$ has a quadratic dependence on the EDSR coupling $D$. When the relaxation time increases, the operation time $\tau_{\sqrt{\text{SWAP}}}$ increases by same magnitude, see Fig. \ref{fig:SWAP}. (b). When the magnitude of strain is weak, the fidelity $F_{\sqrt{\text{SWAP}}}$ is enhanced by asymmetric strain, which is different from single-qubit gates. Except for those decoherence sources for single-qubit operation, the decoherence of two-qubit entanglement is actually induced by the variation of $\sqrt{\text{SWAP}}$ coupling $J_{dd}$, in this region of weak strain. Similar to qubit splitting $\hbar\omega$, the coupling $J_{dd}$ is influenced by charge noise, see Supplementary \cite{supp}. When the strain is small, the pure dephasing due to $J_{dd}$ could dominate the decoherence. However, when strain is more negative ($|\epsilon_{xx}|, |\epsilon_{yy}|>0.05\%$), the main decoherence is still from relaxation due to phonon. In conclusion, a technical route to construct long-range, high-fidelity, and highly feasible two-qubit gates is proposed.

\section{Discussion and Conclusion}

The tunable strain provides a knob for controlling the electric manipulation of the boron-based heavy hole spin qubit and better feasibility compared with the previous work \cite{Salfi2016Nano,Salfi2016PRL}. One may further optimize the operation performance through variation of depth of the acceptor atom, choice of materials, and magnetic field orientation. Moreover, the flexible tunability of the strain is not limited to heavy-hole bound to boron in silicon. It is expected that the qubit operation based on light hole spin could also be optimized with tunable strain when the light hole spin states are the ground states. Light hole-based qubit is also promising since it can be manipulated faster \cite{Salfi2016PRL}. However, there are two drawbacks for light hole qubits: (i) the light hole-based qubit is more sensitive to the charge noise, and (ii) the sweet spots of light hole-based qubits require much higher electric fields to produce. According to the results of this paper, by introducing strain, we expect these problems could be solved.

In conclusion, we investigate the effect of tunable strain on the acceptor hole spin qubit. Compared to the vertical electric field, strain provides a way to tune the key quantities: LH-HH splitting and spin-hole coupling of the system independently. With aid of strain, the required electric field for sweet spots can be lowered. And a second-order sweet spot appears, where the qubit coherence is improved. At sweet spots, the relaxation, which dominates the qubit decoherence, can be suppressed by tuning strain, and quality factor $Q$ can be enhanced accordingly. A concrete parameter regime is specified for high-fidelity quantum gates of hole spin qubits. For the strain $-0.07 \%<\epsilon_{xx} \approx \epsilon_{yy} <-0.05 \%$, and $d=25$ nm, the fidelity above 99.99 \% for single-qubit operation and 99.9 \% for two-qubit operation can be achieved. Thus, all-electric qubit operations can be constructed with fidelity well beyond the fault-tolerant threshold of quantum computing, at a low electric field, and with a large separation between the qubits, which is crucial for scaling up quantum processors. The proposed scheme of the boron-based spin qubit with tunable strain could pave a way for building a large-scale fault-tolerant spin-based quantum computer.

\begin{acknowledgments}
This work is supported by the National Natural Science Foundation of China (Grant Nos. 11904157, 62174076, 92165210), Shenzhen Science and Technology Program (No. KQTD20200820113010023), and Guangdong Provincial Key Laboratory (No. 2019B121203002).
\end{acknowledgments}

\begin{widetext}
\appendix
\setcounter{figure}{0}
\setcounter{equation}{0}
\section*{Supplementary information}
 \section{Strain}\label{sec:strain}
Interaction of holes with strain plays an important role. The interaction is described by the Hamiltonian \cite{Bir19631}:
\begin{equation}
H_{\epsilon}=a'Tr[\epsilon]+b'((J_{x}^{2}-\frac{5}{4}I)\epsilon_{xx}+c.p.)+(2d'/\sqrt{3})(\{J_{x},J_{y}\}\epsilon_{xy}+c.p.),
\end{equation}
where c.p. means cyclic permutations, $a'$, $b'$ and $d'$ are deformation potential, $\{J_{x},J_{y}\}=1/2(J_{x}J_{y}+J_{y}J_{x})$ is anti-commutator. The matrix form of it in the basis of hole spin states $\{\Psi_{+3/2}$, $\Psi_{-3/2}$, $\Psi_{+1/2}$, $\Psi_{-1/2}\}$:
{\tiny
\begin{equation}\label{Eq:strain1}
\begin{aligned}
H_{\epsilon}&=a'\left(\begin{array}{cccc}
\epsilon & 0 & 0 & 0 \\
0 & \epsilon & 0 & 0 \\
0 & 0 & \epsilon & 0 \\
0 &0 & 0 & \epsilon
\end{array}\right)
+b'\left(\begin{array}{cccc}
-\dfrac{1}{2}(\epsilon_{xx}+\epsilon_{yy})+\epsilon_{zz} & 0 &0&  \dfrac{\sqrt{3}}{2}(\epsilon_{xx}-\epsilon_{yy})  \\
0 & -\dfrac{1}{2}(\epsilon_{xx}+\epsilon_{yy})+\epsilon_{zz} &  \dfrac{\sqrt{3}}{2}(\epsilon_{xx}-\epsilon_{yy})  & 0 \\
0 &  \dfrac{\sqrt{3}}{2}(\epsilon_{xx}-\epsilon_{yy})  & \dfrac{1}{2}(\epsilon_{xx}+\epsilon_{yy})-\epsilon_{zz} & 0 \\
 \dfrac{\sqrt{3}}{2}(\epsilon_{xx}-\epsilon_{yy})  &0 & 0 &  \dfrac{1}{2}(\epsilon_{xx}+\epsilon_{yy})-\epsilon_{zz}
\end{array}\right)\\
&+\dfrac{d'}{\sqrt{3}}\left(\begin{array}{cccc}
0 & 0 & \sqrt{3}(\epsilon_{xz}-i\epsilon_{yz}) & -i\sqrt{3}\epsilon_{xy} \\
0 & 0 & i\sqrt{3}\epsilon_{xy} & -\sqrt{3}(\epsilon_{xz}+i\epsilon_{yz}) \\
\sqrt{3}(\epsilon_{xz}+i\epsilon_{yz}) & -i\sqrt{3}\epsilon_{xy}  & 0 & 0\\
i\sqrt{3}\epsilon_{xy}  &-\sqrt{3}(\epsilon_{xz}-i\epsilon_{yz}) & 0& 0
\end{array}\right).
\end{aligned}
\end{equation}
}
The Hamiltonian can be divided into three parts respectively corresponding to $a'$, $b'$, and $d'$, the three Bir-Pikus deformation potentials. The first part is trivial in the form of the identity matrix. And in the second part, there are two kinds of terms. The diagonal terms split heavy hole and light hole. And the off-diagonal terms rotate spin and hole state simultaneously, which is similar to $pE_{z}$-term from $H_{T_{d}}$. These terms appear only if $\epsilon_{xx}$ and $\epsilon_{yy}$ are different, which means linear strain in the x-y plane is asymmetric. The form of the last part is similar to that of $H_{T_{d}}$. Again, there are similar terms for rotating spin and hole state simultaneously. And the rest of the terms change hole state the with orientation of spin reserved.

\begin{figure*}[htbp]
\centering
\includegraphics[width=0.8\linewidth]{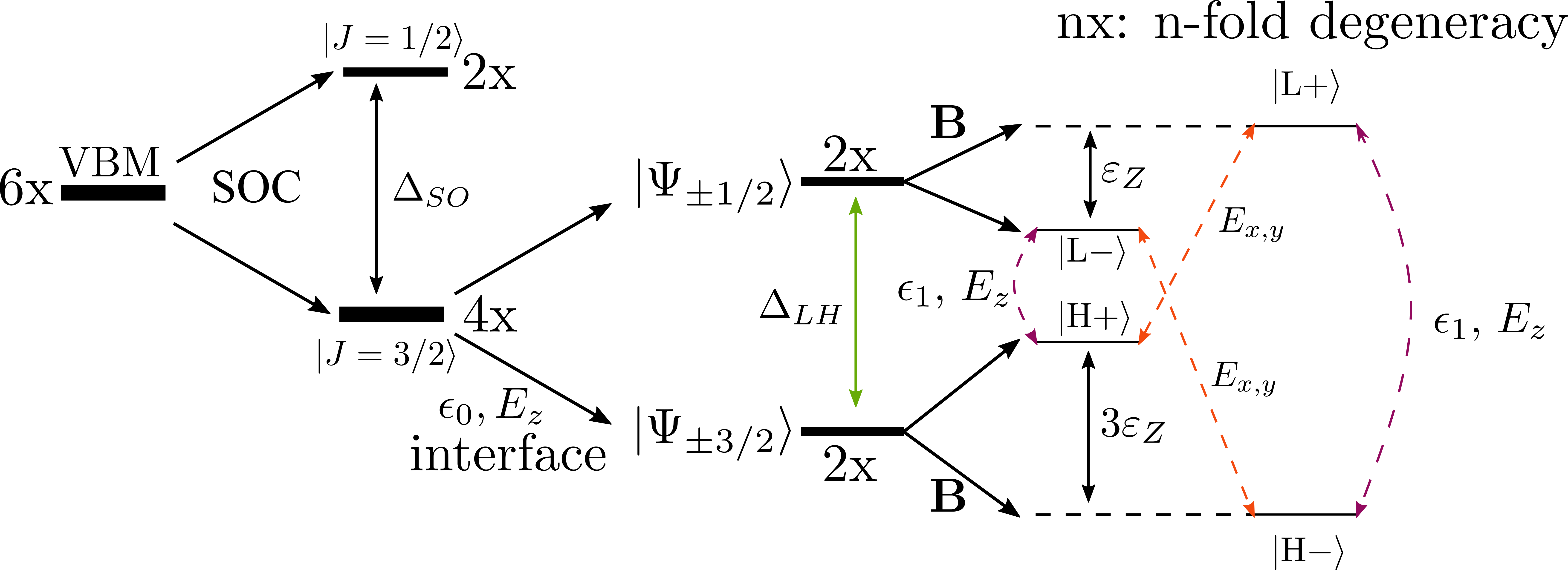}
\caption{An extended version of Fig. 1 (c) in the main text. Schematic of energy levels of hole spin states. Due to the spin-orbit coupling, states on the six-fold degenerate valence band maximum (VBM) are split into a two-fold degenerate state named split-off band and a four-fold degenerate state. Then the four-fold degenerate states can be divided into the heavy hole and light hole states, by strain or electric field. The magnetic field further splits the 2-fold degeneracies of the spin states of the heavy hole and light hole. In this work, the heavy hole spin state is always grounded. The qubit states are spin-hole mixed states due to interaction with asymmetric strain and $T_{d}$-interaction with electric field $E_{z}$. The LH-HH transition due to the in-plane electric field plays an important role in EDSR.}
\label{fig:device0-e}
\end{figure*}

\section{EDSR coupling}\label{sec: EDSR}

From Sec. \ref{sec: operation} of the main text, the EDSR coupling D is:
\begin{equation}
	D=p_{eff} C_{1}= |t|p_{eff} \frac{|a-c|}{N_{1}N_{2}}.
\end{equation}\label{Eq:edsr}
The form here can be transformed to a more practical form with realistic parameters:
\begin{equation}\label{eq:D}
	D=|\frac{|t|p_{eff} (-4\varepsilon_{Z}+\sqrt{(\Delta-2\varepsilon_{Z})^2+4|t|^2}-\sqrt{(\Delta+2\varepsilon_{Z})^2+4|t|^2}) }{2\sqrt{[|t|^{2}+(\frac{\Delta-2\varepsilon_{Z}+\sqrt{(\Delta-2\varepsilon_{Z})^2+4|t|^2}}{2})^2][t^{2}+(\frac{\Delta+2\varepsilon_{Z}+\sqrt{(\Delta+2\varepsilon_{Z})^2+4|t|^2}}{2})^2]}}|,
\end{equation}
where the $\Delta$ is the LH-HH splitting, $t$ is SHC, $\epsilon_{Z}$ is the Zeeman energy, and $p_{eff}$ is the effective LH-HH transition. The dependence of the EDSR coupling $D$ on the magnetic field $B$ is plotted in Fig. \ref{fig:T1}. (a). The EDSR coupling is almost linearly dependent on the magnetic field. Later on, we show that the relaxation rate is more sensitive to the magnetic field, showed in Fig. \ref{fig:T1}. (b). Thus, the quality factor can be enhanced by a smaller magnetic field. High fidelity of the qubit operations can be achieved.

\section{Dephasing due to the charge noise} 
All-electrical manipulation of spin qubits can be controlled easily and is beneficial to the integration of qubits. However, it is unavoidable to bring decoherence from charge noise by the same path. Defects in the crystal induce unexpected electric field $\mathcal{F}_{z}$, which is the main source of charge noise. The dephasing rate is $1/T_{\phi}=V^{2}\tau/(2\hbar^{2})$ \cite{Bermeister2014}. To calculate the energy fluctuation $V$ of the unexpected electric field on qubit splitting, the derivative of the qubit splitting relative to the electric field is needed:
\begin{equation}
V=\dfrac{\partial \hbar\omega}{\partial E_{z}}E_{z}+\dfrac{\partial^{2}\hbar\omega}{\partial E_{z}^{2}}E_{z}^{2}.
\end{equation}

\begin{figure}[htbp]
\centering
\includegraphics[width=1\textwidth]{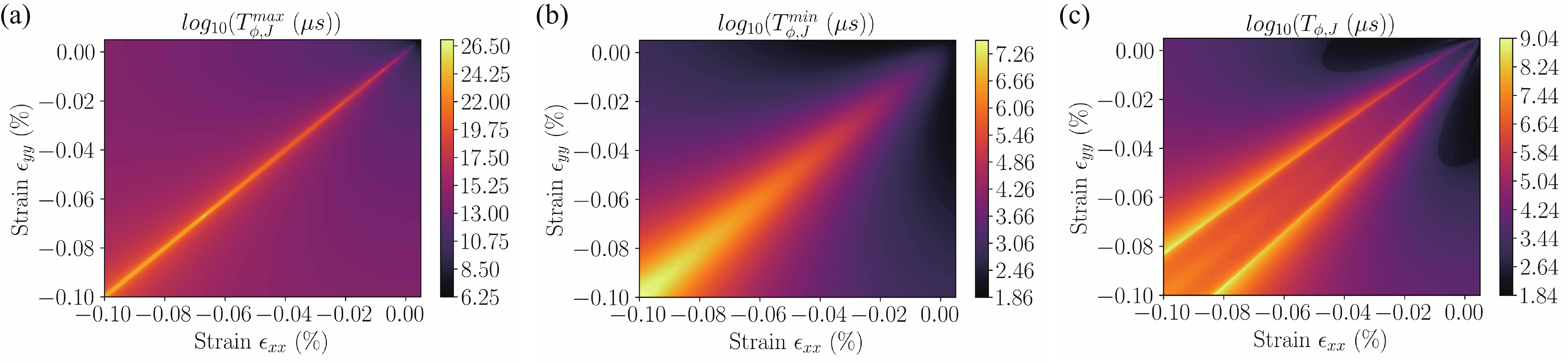}%
\caption{Dephasing of spin qubits from $J_{dd}$. $B=0.5$ T, $E_{ac}=10^{4}$ V/m, $d=4.6$ nm and $R_{12}=25$ nm. The window of the vertical electric field is set as $[1,20]$ MV/m. (a) The max value of the dephasing time $log_{10}(T_{\phi,J}^{max})$. (b) The minimum value of the dephasing time $log_{10}(T_{\phi,J}^{min})$. Except for the sweet spot, Dephasing from $J_{dd}$ is weaker than that from qubit splitting $\hbar\omega$. (c) The dephasing time due to the charge noise $T_{2,J}$ corresponds to $Q_{\sqrt{\text{SWAP}}}$.}
\label{fig:T2S}
\end{figure} 

Qubit splitting is dependent on the Zeeman energy, LH-HH splitting $\Delta_{LH}$, and spin-hole coupling $t$. Except for Zeeman energy and terms from interaction with strain, all of them are relative to the electric field. Thus, it is hard to show a compacted form of $\frac{\partial \hbar\omega}{\partial E_{z}}$ and $\frac{\partial^{2}\hbar\omega}{\partial E_{z}^{2}}$ here. However, we found that the dephasing due to the second-order electrical noise is much weaker. The first-order electrical noise plays an important role in decoherence. Moreover, for the $\sqrt{\text{SWAP}}$ gate realized by dipole-dipole interaction, the coupling strength $J_{dd}$ is also fluctuated by charge noise. And the dephasing rate can be obtained similarly. The dephasing due to the charge noise on $J_{dd}$ is weaker compared to that on qubit splitting out of the sweet spot.

\section{Relaxation due to the interaction with phonon}\label{sec: t1}
When the hole spin qubit is operated at the sweet spot, the relaxation may lead to the decoherence of the qubit. In this work, relaxation is obtained by considering the interaction of holes with phonons. The relaxation rate is calculated by the Fermi's golden rule:
\begin{equation}
\dfrac{1}{T_{n\rightarrow n'}}=\dfrac{2\pi}{\hbar}|\langle n'|H'|n\rangle|^{2}\rho(E_{n'}).
\end{equation}
The relaxation is due to the interaction with phonon. Thus, perturbation $H'$ is hole-phonon interaction. Fermi's golden rule transforms to:
\begin{equation}\label{eq:fg}
\dfrac{1}{T_{n\rightarrow n'}}=\dfrac{2\pi}{\hbar}\sum_{i,j,s,\mathbf{q}_{s}}|\langle n',n_{\mathbf{q}}+1|H_{\epsilon ijs}|n,n_{\mathbf{q}}\rangle|^{2}\times \delta(E_{n}-E_{n}'-\hbar\omega_{qs}).
\end{equation}
The transition from $|n\rangle$ to $|n'\rangle$ is complicated by emission of phonon. The energy of phonon is $\hbar\omega_{qs}=\hbar v_{s}q_{s}$. The sum in the formula is the density of states with proper energy. $\mathbf{q}_{s}$ is the phonon wave vector. And $s=l,t_{1},t_{2}$ are the phonon polarization. $\sum_{i,j}H_{\epsilon}=\sum_{i,j}D_{ij}\epsilon_{ijs}$ is the electron-phonon interaction, where $D_{ij}$ are deformation potential matrices from Bir-Pikus Hamiltonian. After some simplification and calculation steps \cite{Srivastava1990,Ehrenreich1956,Salfi2016Nano}, the relaxation rate is: 
\begin{equation}\label{eq:t1}
\frac{1}{T_{1}}=\frac{(\hbar \omega)^{3}}{20 \hbar^{4} \pi \rho}\left[\sum_{i}\left|\left\langle n^{\prime}\left|D_{i i}\right| n\right\rangle\right|^{2}\left(\frac{2}{v_{l}^{5}}+\frac{4}{3 v_{t}^{5}}\right)+\sum_{i \neq j}\left|\left\langle n^{\prime}\left|D_{i j}\right| n\right\rangle\right|^{2}\left(\frac{2}{3 v_{l}^{5}}+\frac{1}{v_{t}^{5}}\right)\right].
\end{equation}

\begin{figure}
\centering
\includegraphics[width=1\textwidth]{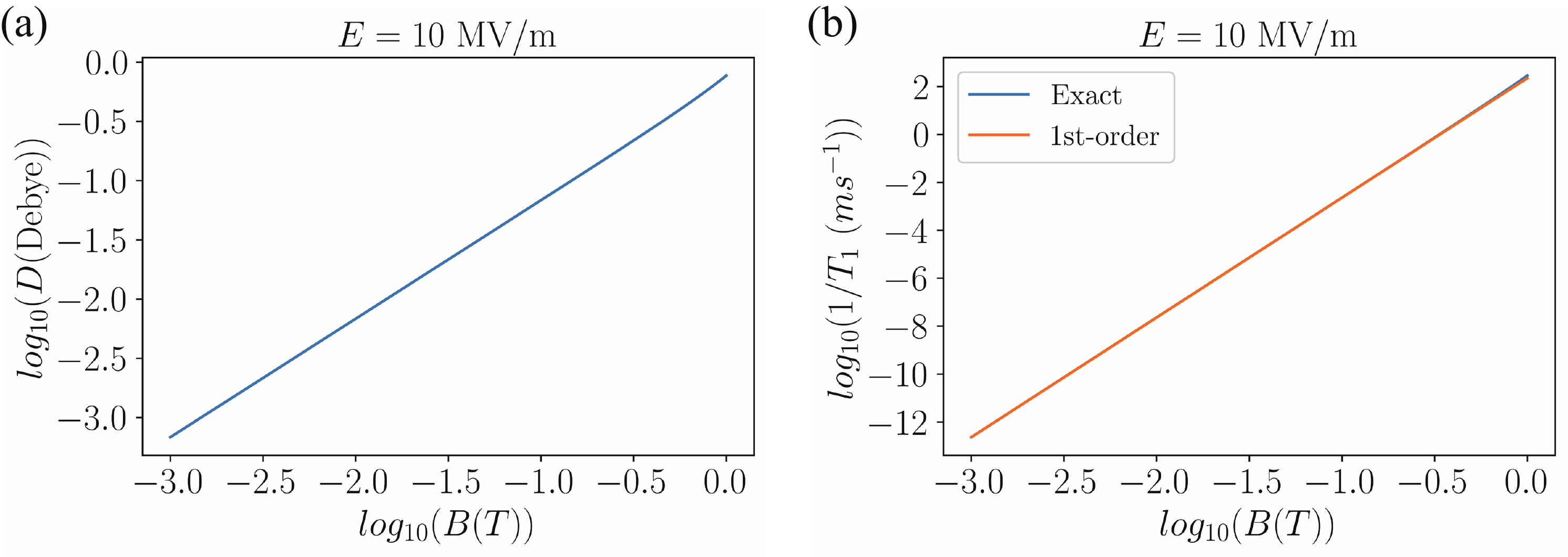}%
\caption{EDSR coupling $D$ and relaxation rate $1/T_{1}$ due to interaction with phonon plotted as functions of magnetic field B. Here, we take the logarithm to the base 10 of them to clarify the difference. (a) The EDSR coupling $D$ is almost linearly dependent on the magnetic field $B$.  (b) The blue (orange) line is calculated in the exact (perturbed) way. They have similar dependence on the magnetic field. As the magnetic field increases, relaxation from exact calculation becomes higher. In conclusion, the quality factor $Q$ can be enhanced with the weaker magnetic field $B$.}
\label{fig:T1}
\end{figure} 

Then, the relaxation time of a hole spin qubit is calculated. For the system, $\left\langle n^{\prime}\left|D_{i j}\right| n\right\rangle$ is $\left\langle g-\left|D_{i j}\right| g+\right\rangle$, which is obtained by Schrieffer-Wolff transformation of $ \tilde{H}=\tilde{H}_{op}+\tilde{H}_{\epsilon ijs}$ ($\tilde{H}=U^{\dagger}_{t0} H U_{t0}$). Thus, for heavy hole qubit:
\begin{equation}
\langle g-|D_{i j}|g+\rangle=\dfrac{1}{E_{g}-E_{e}}(H'_{g-,e-}H'_{e-,g+}+H'_{g-,e+}H'_{e+,g+}).
\end{equation}
And the deformation potential:
\begin{equation}
\begin{aligned}
D_{ii}&=b'(J_{i}^{2}-\dfrac{5}{4}),
\\D_{i j}&=\dfrac{2d'}{\sqrt{3}}\{J_{i},J_{j}\} ,(i\neq j).
\end{aligned}
\end{equation}
For heavy hole qubit:
\begin{equation}
\begin{aligned}
|\langle g-|D_{xx}|g+\rangle|^{2}&=|\langle g-|D_{yy}|g+\rangle|^{2}=|\langle g-|D_{zz}|g+\rangle|^{2}=|\langle g-|D_{xy}|g+\rangle|^{2}=0,\\
|\langle g-|D_{yz}|g+\rangle|^{2}&=|\langle g-|D_{zx}|g+\rangle|^{2}=16d^{'2}(\dfrac{\tilde{\varepsilon}_{Z}}{\tilde{\Delta}})^{2}.\\
\end{aligned}
\end{equation}
where $\tilde{\varepsilon}_{Z}=\varepsilon_{Z}t/\tilde{\Delta}$, $\tilde{\Delta}=\sqrt{\Delta^{2}+t^{2}}$. Phonon-induced relaxation of heavy hole spin qubit:
\begin{equation}
\frac{1}{T_{1}}=\frac{(\hbar \omega)^{3}}{20 \hbar^{4} \pi \rho}(\dfrac{\tilde{\varepsilon}_{Z}}{\tilde{\Delta}})^{2}\left[32d^{'2}\left(\frac{2}{3 v_{l}^{5}}+\frac{1}{v_{t}^{5}}\right)\right].
\end{equation}

According to previous paragraphs, $H_{Z}$ should be treated as a perturbation to induce relaxation. However, $H_{Z}$ can be exactly treated to obtain the relaxation, which is more precise than that from the perturbation step. The main difference between the exact and perturbed way is deformation potential $\langle g-|D_{ij}|g+\rangle$. Now, Zeeman Hamiltonian $H_{Z}$ is considered exactly, which means $H_{hp}$ is projected into the exact form of acceptor qubit. The deformation potential can be obtained without using the Schrieffer-Wolff transformation:
\begin{equation}
\begin{aligned}
&|\langle g-|D_{xx}|g+\rangle|^{2}=|\langle g-|D_{yy}|g+\rangle|^{2}=|\langle g-|D_{zz}|g+\rangle|^{2}=|\langle g-|D_{xy}|g+\rangle|^{2}=0,\\
&|\langle g-|D_{yz}|g+\rangle|^{2}=|\langle g-|D_{zx}|g+\rangle|^{2}=d^{'2}(\dfrac{|t|(a-c)}{N_{1}N_{2}})^{2}.\\
\end{aligned}
\end{equation}
Then, substituting them into Eq. (\ref{eq:t1}), relaxation of heavy hole acceptor qubit:
\begin{equation}
\frac{1}{T_{1}}=\frac{(\hbar \omega)^{3}}{20 \hbar^{4} \pi \rho}[\dfrac{|t|(a-c)}{N_{1}N_{2}}]^{2}\left[2d^{'2}\left(\frac{2}{3 v_{l}^{5}}+\frac{1}{v_{t}^{5}}\right)\right].
\end{equation}
To verify the validity of the exact method, relaxation from the exact or perturbed way is compared in Fig. \ref{fig:T1}. They are almost consistent when the magnetic field is weak. That means the perturbed result matches up to the first-order part of the exact result. It is reasonable to estimate the relaxation due to phonon in the exact way introduced in this subsection.

\begin{figure}
\centering
\includegraphics[width=1\textwidth]{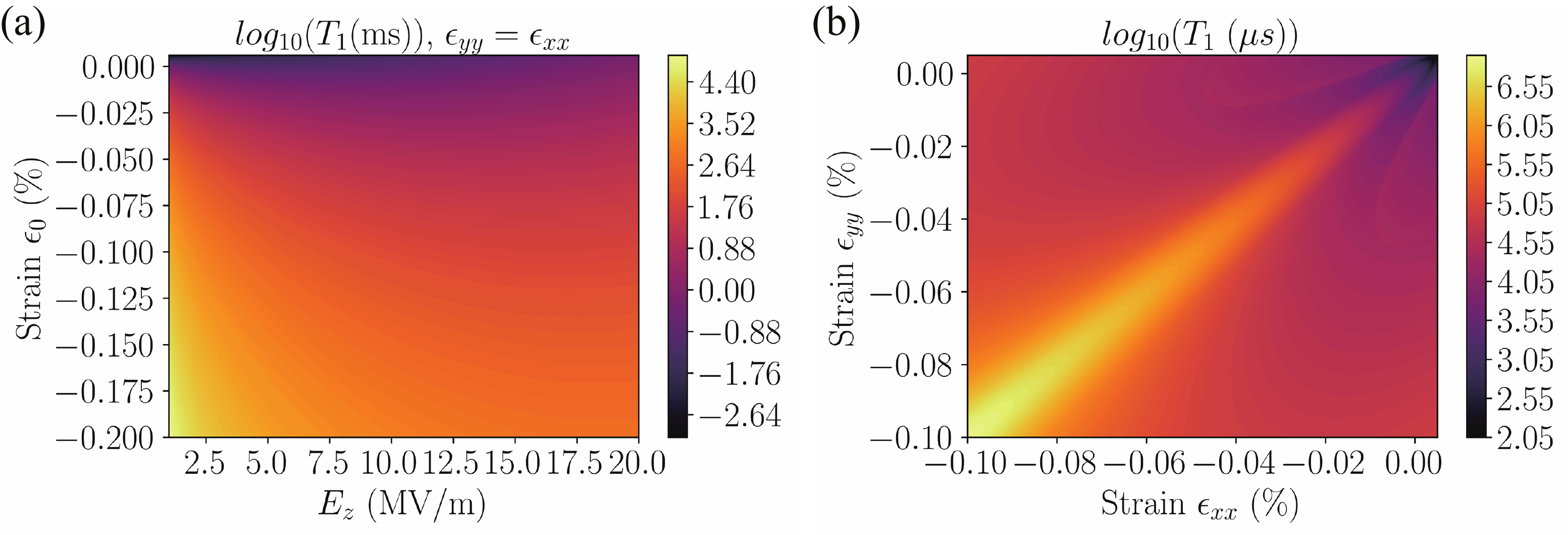}
\caption{The relaxation rate due to interaction with phonon. $B=0.5$ T, $E_{ac}=10^{4}$ V/m, $d=4.6$ nm. Here, we take the logarithm to the base 10 of them to clarify the difference. (a) $T_{1}$ is plotted as function of symmetric strain $\epsilon_{0}$ and vertical electric field $E_{z}$. The relaxation time is longer with stronger strain. (b) $T_{1}$ is plotted as function of strain $\epsilon_{xx}$ and $\epsilon_{yy}$.}
\label{fig:T1s}
\end{figure} 

\section{Tolerance to the electric field at sweet spot}\label{sec:FWHM}
With aid of asymmetric strain, there can be two sweet spots for the single-qubit operation of the acceptor-based qubit. And as the strain increases, the two sweet spots can merge together and form a second-order sweet spot. Around the region of the second-order sweet spot, the qubits are still insensitive to the charge noise. The coherence performance of the qubits is preserved. In Fig. \ref{fig:FWHM}, the $E_{tol}$ is defined as full width at half maximum of $E_{max}$. Around most of the high-Q area, the $E_{tot}$ exceeds 1000 V/m.

\begin{figure*}[!htbp]
\centering
\includegraphics[width=0.5\textwidth]{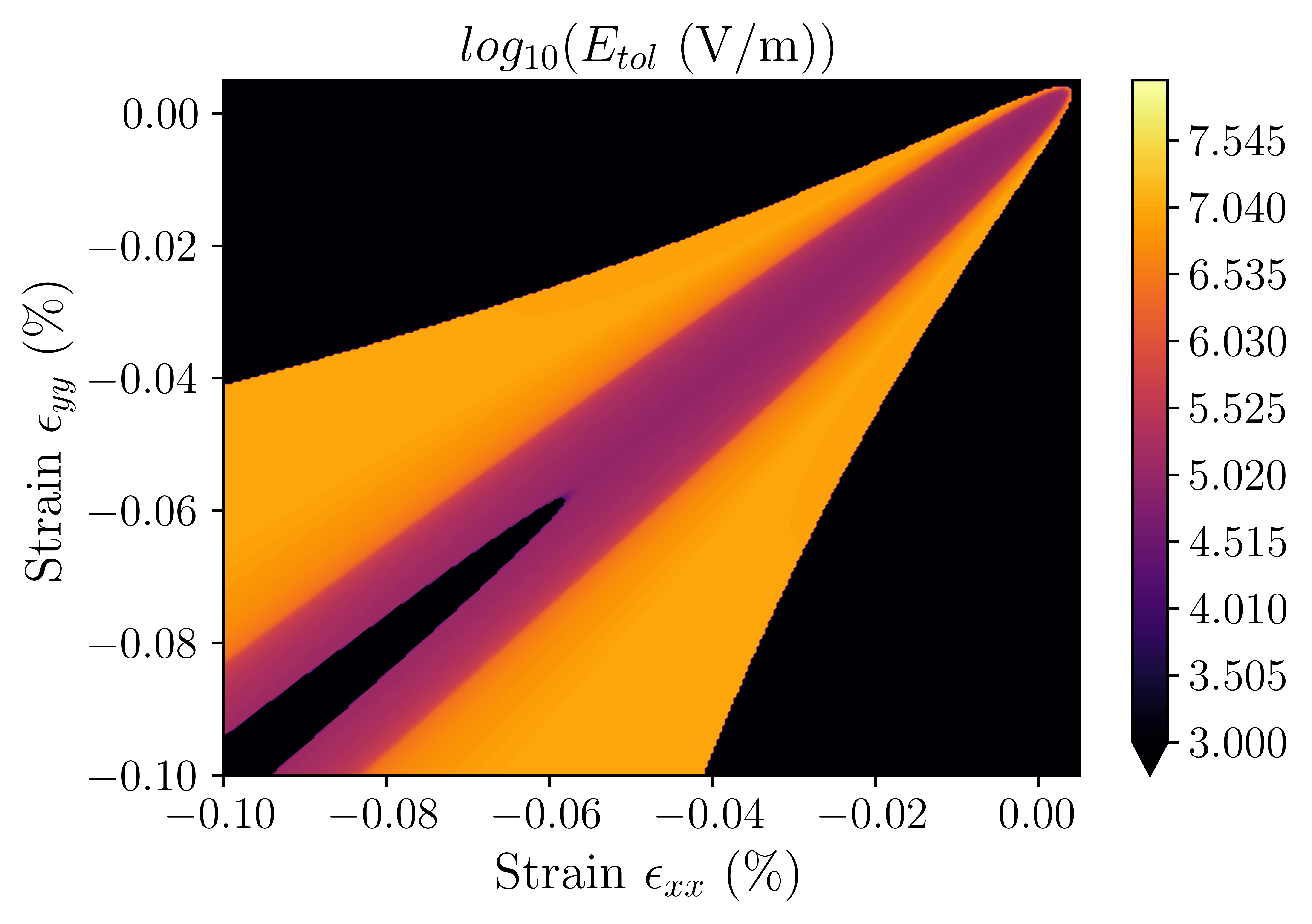}%
\caption{Tolerance to the electric field $E_{tol}$. $B=0.5$ T, $E_{ac}=10^{4}$ V/m, $d=4.6$ nm. The window of the vertical electric field is set as $[1,20]$ MV/m. The figures are plotted as functions of strain $\epsilon_{xx}$ and $\epsilon_{yy}$.}
\label{fig:FWHM}
\end{figure*}

\section{Two qubit gate: dipole-dipole interaction}\label{section: dd}
The two-qubit gate might be realized by the electric dipole of acceptors and spin-orbit coupling. To import this effect, we assume two acceptors are separated by the distance of R, which can make exchange interaction ignored. And the interaction between the acceptors is Coulomb interaction $V_{12}$. Hamiltonian of the two acceptor qubits is:
\begin{equation}\label{Eq:dd}
H=H_{op}^{1}+H_{op}^{2}+V_{12},
\end{equation} 
where $H_{op}^{1(2)}$ is the operation Hamiltonian of acceptor 1 (2). Two qubits subspace is contained in a 16$\times$16 product subspace $\{|mn\rangle\}=\{|m^{1}\rangle\}\otimes\{|n^{2}\rangle\}$. $m, n\in \{g+,g-,e+,e-\}$. In the subspace of the single qubit, the Hamiltonian is obvious:
\begin{equation}
\langle m(n)|H_{op}^{1(2)}|m'(n')\rangle=
\begin{pmatrix}
\varepsilon_{g+} & 0 & 0 & 0 \\
0 & \varepsilon_{g-} & 0 & 0 \\
0 & 0 & \varepsilon_{e+} & 0 \\
0 & 0 & 0 & \varepsilon_{e-}
\end{pmatrix}.
\end{equation}
Hamiltonian of the Coulomb interaction is:
\begin{equation}
\langle mn|V_{12}|m'n'\rangle=\int dr_{1}^{3}dr_{2}^{3}\dfrac{e^{2} \Psi_{m}^{\dagger}(\mathbf{r}_{1}) \Psi_{n}^{\dagger}(\mathbf{r}_{2}) \Psi_{m'}(\mathbf{r}_{1}) \Psi_{n'}(\mathbf{r}_{2})}{4\pi \epsilon |\mathbf{r}_{1}\mathbf{r}_{2}|}.
\end{equation}
And we suppose the separation $\mathbf{R}_{12}=\mathbf{r}_{1}-\mathbf{r}_{1}$ between the acceptors is larger than the dipole moment of the acceptors. Then the Coulomb interaction can be expanded by multi-pole approximation. The lowest-order non-zero term is:
\begin{equation}
\langle mn|V_{12}|m'n'\rangle=\dfrac{e^{2}}{4\pi\epsilon R_{12}^{5}} [R^{2}_{12}\langle \delta\mathbf{r}_{1}\rangle_{nn'} \langle \delta\mathbf{r}_{2}\rangle_{mm'}-3(\langle \delta\mathbf{r}_{1}\rangle_{nn'}\cdot\mathbf{R}_{12})(\langle \delta\mathbf{r}_{2}\rangle_{mm'}\cdot\mathbf{R}_{12})],
\end{equation}
where
\begin{equation}
\langle \delta\mathbf{r}_{1}\rangle_{nn'}=\int dr_{i}^{3}(\mathbf{r}_{i}-\mathbf{r}_{i})\Psi_{n}^{\dagger}(\mathbf{r}_{i})\Psi_{n'}(\mathbf{r}_{i}).
\end{equation}
Defining $\mathbf{r}'_{i}=\mathbf{r}_{i}-\mathbf{R}_{i}$ the hole coordinate relative to the ion, and assuming the relative position in the x-y plane: $\mathbf{R}=R\cos(\theta_{E})\hat{x}+R\sin(\theta_{E})\hat{y}$, the Coulomb interaction:
{\tiny
\begin{equation}
\langle mn|V_{12}|m'n'\rangle=\dfrac{(1-3\cos^{2}(\theta_{E}))\langle m|ex_{1}'|m'\rangle \cdot\langle n|ex_{2}'|n'\rangle+ (1-3\sin^{2}(\theta_{E}))\langle m|ey_{1}'|m'\rangle \cdot\langle n|ey_{2}'|n'\rangle+\langle m|ez_{1}'|m'\rangle \cdot\langle n|ez_{2}'|n'\rangle}{4\pi\epsilon R^{3}}.
\end{equation}}

The dipole matrix required by the Coulomb interaction:
\begin{equation}
\langle m(n)|ex'|m'(n')\rangle=\alpha
\begin{pmatrix}
0 & C_{1} & C_{2} & 0 \\
C_{1}^{*} & 0 & 0 & -C_{2} \\
C_{2} & 0 & 0 & C_{1} \\
0 & -C_{2} & C_{1}^{*} & 0
\end{pmatrix},
\end{equation}
and
\begin{equation}
\langle m(n)|ey'|m'(n')\rangle=\alpha
\begin{pmatrix}
0 & -iC_{1} & -iC_{2} & 0 \\
iC_{1}^{*} & 0 & 0 & iC_{2} \\
iC_{2} & 0 & 0 & -iC_{1} \\
0 & -iC_{2} & iC_{1}^{*} & 0
\end{pmatrix}.
\end{equation}
Substituting them into the Eq. (\ref{Eq:dd}), and projecting into two-qubit subspace $\{|g-g-\rangle,|g-g+\rangle,|g+g-\rangle,|g+g+\rangle\}$:
\begin{equation}
\langle mn| H| m'n'\rangle=
\begin{pmatrix}
\varepsilon^{a}_{g-}+\varepsilon^{b}_{g-} & 0_{1} & 0 & 3d_{11}e^{-2i\theta_{E}} \\
0 & \varepsilon^{a}_{g-}+\varepsilon^{b}_{g+} & -d_{11} & 0 \\
0 & -d_{11} & \varepsilon^{a}_{g+}+\varepsilon^{b}_{g-} & 0 \\
3d_{11}e^{2i\theta_{E}} &0 & 0 & \varepsilon^{a}_{g+}+\varepsilon^{b}_{g+}
\end{pmatrix},
\end{equation}
where a,b corresponding to acceptor 1,2, and $d_{11}$ is defined as $d_{11}=\dfrac{\alpha^{2} C_{1}^{a}C_{1}^{b}}{4\pi\epsilon R_{12}^{3}}$. Actually, $d_{11}$ can be approximated to:
\begin{equation}
d_{11}\approx \dfrac{D^{2}}{4\pi\epsilon R_{12}^{3}}.
\end{equation} 
From the above, we know that form of entanglement of acceptor qubits is determined by relative separation between acceptors. However, entanglement between $|g-g+\rangle$ and $|g+g-\rangle$ is independent on that. $\sqrt{\text{SWAP}}$ gate can be constructed by assuming $\varepsilon^{a}_{g-}+\varepsilon^{b}_{g+}=\varepsilon^{a}_{g+}+\varepsilon^{b}_{g-}$.

\end{widetext}

\providecommand{\noopsort}[1]{}\providecommand{\singleletter}[1]{#1}%

\end{document}